\documentclass[12pt]{article}
\usepackage{amsmath}
\usepackage{graphicx}
\usepackage{enumerate}
\usepackage{natbib}
\usepackage{url} 
\usepackage{amsmath, amsthm}
\usepackage{amssymb}
\usepackage[margin=0.9in, includefoot]{geometry}
\usepackage[singlespacing]{setspace}
\usepackage{enumitem}
\usepackage{comment}
\usepackage{hyperref}
\usepackage{sectsty}
\sectionfont{\centering}
\usepackage{secdot}
\usepackage{authblk}

\newcommand{\blind}{1}

\begin{document}

\newtheorem{prop}{Proposition}
\numberwithin{prop}{section}
\newtheorem{theorem}{Theorem}
\numberwithin{theorem}{section} 

\newtheorem{lemma}{Lemma}
\newtheorem{corollary}{Corollary}
\numberwithin{corollary}{section}
\newtheorem{remark}[theorem]{Remark}


\if1\blind
{
  \title{\Large\bf Group Inverse-Gamma Gamma Shrinkage for Sparse Regression with Block-Correlated Predictors}
  \author[1]{\normalsize Jonathan Boss\thanks{Corresponding Author: {\it bossjona@umich.edu}}}
  \author[2]{Jyotishka Datta}
  \author[3]{Xin Wang}
  \author[3]{Sung Kyun Park}
  \author[1]{Jian Kang}
  \author[1]{Bhramar Mukherjee}
  \affil[1]{Department of Biostatistics, University of Michigan}
  \affil[2]{Department of Statistics, Virginia Polytechnic Institute and State University}
  \affil[3]{Department of Epidemiology, University of Michigan}
  \date{}
  \maketitle
} \fi

\if0\blind
{
  \bigskip
  \bigskip
  \bigskip
  \begin{center}
    {\LARGE\bf Group Inverse-Gamma Gamma Shrinkage for Sparse Regression with Block-Correlated Predictors}
\end{center}
  \medskip
} \fi

\vspace{-4 mm}
\begin{abstract}
Heavy-tailed continuous shrinkage priors, such as the horseshoe prior, are widely used for sparse estimation problems. However, there is limited work extending these priors to predictors with grouping structures. Of particular interest in this article, is regression coefficient estimation where pockets of high collinearity in the covariate space are contained within known covariate groupings. To assuage variance inflation due to multicollinearity we propose the group inverse-gamma gamma (GIGG) prior, a heavy-tailed prior that can trade-off between local and group shrinkage in a data adaptive fashion. A special case of the GIGG prior is the group horseshoe prior, whose shrinkage profile is correlated within-group such that the regression coefficients marginally have exact horseshoe regularization. We show posterior consistency for regression coefficients in linear regression models and posterior concentration results for mean parameters in sparse normal means models. The full conditional distributions corresponding to GIGG regression can be derived in closed form, leading to straightforward posterior computation. We show that GIGG regression results in low mean-squared error across a wide range of correlation structures and within-group signal densities via simulation. We apply GIGG regression to data from the National Health and Nutrition Examination Survey for associating environmental exposures with liver functionality.
\end{abstract}

\noindent%
\vfill

\newpage
\section{INTRODUCTION}
\label{sec:intro}

\begin{figure}[!ht]
    \centering
    \includegraphics[scale=1.0, height = 0.65\textheight, width = 0.8\linewidth]{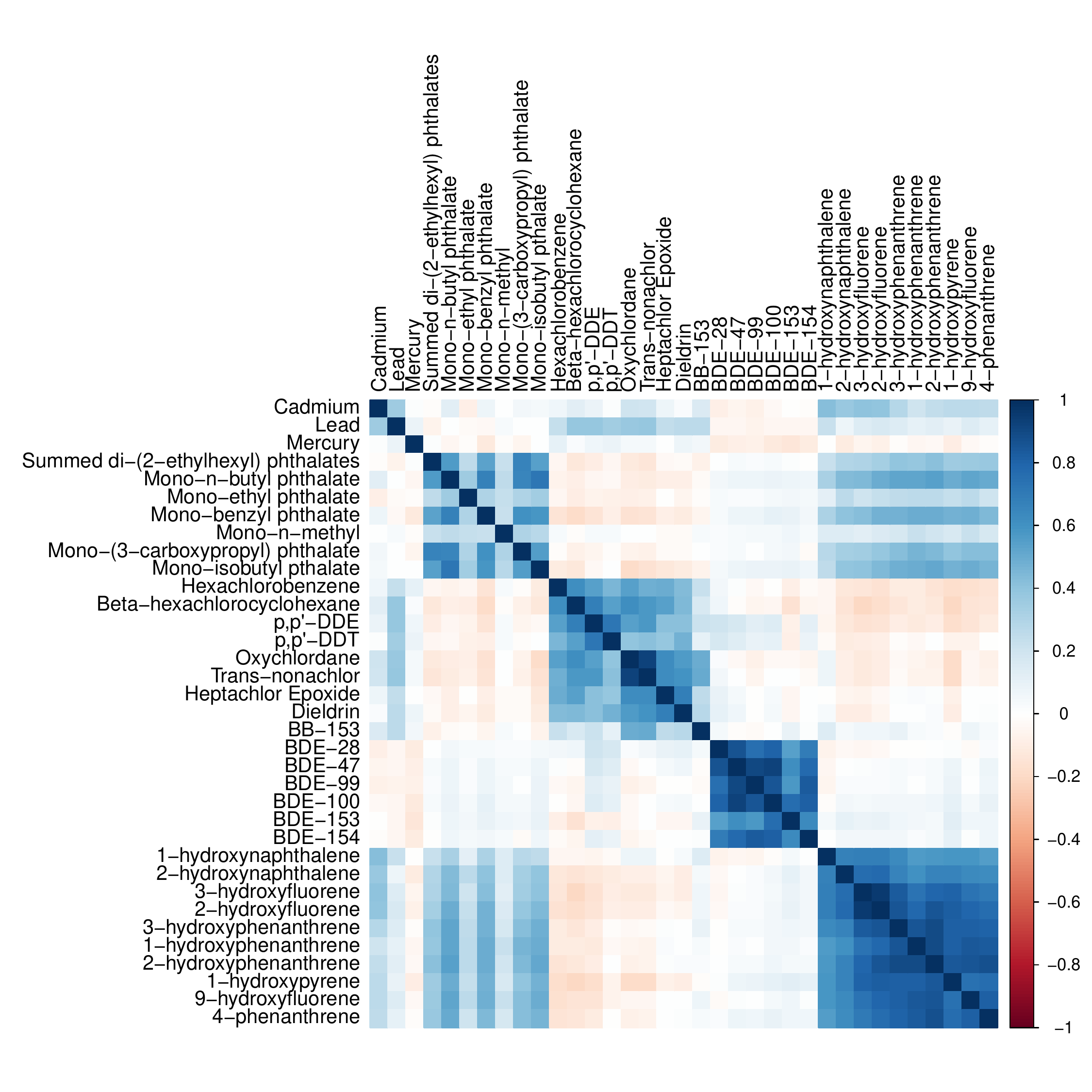}
    \caption{Pairwise Spearman correlation plot between metals, phthalates, organochlorine pesticides, polybrominated diphenyl ethers, and polycyclic aromatic hydrocarbons from the 2003-2004 National Health and Nutition Examination Survey ($n = 990$).}
    \label{fig:ewas_corrplot}
\end{figure}

Regression with grouped features is a common problem in many biomedical applications. Some examples include metabolomics data, where metabolites are grouped by subpathway membership, neuroimaging data, where adjacent voxels are spatially grouped, and environmental contaminants data, where exposures are grouped by chemical structure, toxicological profile, and pharmacokinetics (see Figure \ref{fig:ewas_corrplot}). In such cases, leveraging relevant grouping information to construct correlated within-group shrinkage profiles may help achieve additional variance reduction beyond comparable methods that ignore the grouping structure. The methodological focus of this article will be on incorporating known covariate grouping information into a continuous shrinkage prior framework.

Ever since the publication of the horseshoe prior \citep{carvalho2009, carvalho2010}, there has been an explosion of continuous shrinkage priors designed for sparse estimation problems, notably generalized double Pareto shrinkage \citep{armagan2013gdp}, Dirichlet--Laplace shrinkage \citep{bhattacharya2015}, horseshoe+ shrinkage \citep{bhadra2017}, and normal beta prime (NBP) shrinkage \citep{bai2019}, among others. These priors have become increasingly  popular for sparse regression problems because of their good theoretical and empirical properties, in addition to their scale mixture representation, which facilitates straightforward and efficient posterior simulation algorithms. The general recipe for constructing a continuous shrinkage prior with good estimation and prediction properties is substantial mass at the origin, to sufficiently shrink null coefficients towards zero, and regularly-varying tails, to avoid overregularizing non-null coefficients \citep{bhadra2016}. Surveying the continuous shrinkage prior literature on regression with known grouping structure, there are many papers which discuss Bayesian group lasso and its applications \citep{kyung2010, li2015, xu2015, hefley2017, kang2019} and several papers which propose extensions to Bayesian sparse group lasso \citep{xu2015}, Bayesian group bridge regularization \citep{mallick2017}, and the Normal Exponential Gamma prior with grouping structure \citep{rockova2014}. \cite{xu2016} introduced the, so called, group horseshoe prior with an emphasis on prediction in Bayesian generalized additive models. However, the group horseshoe prior does not reduce to the horseshoe prior for a group of size one, meaning that the group horseshoe prior, as proposed by \cite{xu2016}, is not a direct generalization of the horseshoe prior.

Bayesian group lasso-style shrinkage is not generally preferred as a default method for estimation problems, as the Laplacian prior has neither an infinite spike at zero nor regularly-varying tails \citep{polsonscott2011, castillo2015, bhadra2016}. The group horseshoe prior of \cite{xu2016} has the desired origin and tail behavior marginally, however no hyperparameter in the prior controls how correlated the shrinkage is within a group. Thus, this prior implicitly assumes that the degree of correlated shrinkage within-group only depends on group size. This assumption is inadequate when we {\it a priori} believe that, irrespective of group size, some groups have more heterogeneous effect sizes than others and, moreover, does not avail the opportunity to learn how correlated the shrinkage should be in a data adaptive manner, which is an intrinsic feature in some application areas. For example, in modeling multiple pollutants, this is a relevant consideration as some exposure classes have more homogeneous toxicological profiles than others \citep{ferguson2014}. From a theoretical perspective, the existing posterior concentration and posterior consistency results for heavy-tailed continuous shrinkage priors all, to our knowledge, apply to independent or exchangeable priors, meaning that there have not yet been any attempts to employ similar arguments for dependent priors.

To address these limitations, we propose the group inverse-gamma gamma (GIGG) prior, which extends the horseshoe and normal beta prime (NBP) priors to incorporate grouping structures. The GIGG prior introduces a group level shrinkage parameter, in addition to the usual global and local shrinkage parameters, such that the induced prior on the product of the group and local shrinkage parameters yields the desired marginal shrinkage profile. This allows the user to control the trade-off between group-level and individual-level shrinkage, leading to relatively low estimation error irrespective of the signal density and the degree of multicollinearity within each group. Additionally, the GIGG prior is constructed such that all parameters have closed-form full conditional distributions, implying that techniques to scale horseshoe regression to large sample sizes and high-dimensional covariate spaces are also applicable to GIGG regression \citep{bhattacharya2016, terenin2019, johndrow2020}. Theoretically, we establish posterior consistency and posterior concentration results for regression coefficients with grouping structure in linear regression models and mean parameters with grouping structure in sparse normal means models with respect to several GIGG hyperparameters and correlation structures. To our knowledge, we are the first to apply existing theoretical frameworks for posterior consistency in the sparse linear regression model \citep{armagan2013} and posterior concentration in the sparse normal means model \citep{datta2013} to a non-exchangeable prior, which will be useful for future evaluations of other non-exchangeable priors.

The structure of the paper is as follows. We start with some theoretical results in Section \ref{sec:theory}, preceded by an intuitive explanation of the GIGG prior in Section \ref{sec:methods}. After the methodological and theoretical discussion, we outline computational details, including hyperparameter estimation via marginal maximum likelihood estimation (MMLE) (Section \ref{sec:computation}). In Section \ref{sec:simulations}, we conduct a simulation study to empirically verify that the intuition and theory developed in Sections \ref{sec:methods} and \ref{sec:theory} hold for linear regression models with group-correlated features. We then apply GIGG regression to data from the 2003-2004 National Health and Nutrition Examination Survey (NHANES) to identify toxicants and metals associated with a biomarker of liver function (Section \ref{sec:ewas}) and conclude with a discussion (Section \ref{sec:discussion}).

\section{METHODS}
\label{sec:methods}

Throughout the article, $N(\boldsymbol{\mu}, \boldsymbol{\Sigma})$ denotes a multivariate normal distribution with mean parameter $\boldsymbol{\mu}$ and variance-covariance matrix $\boldsymbol{\Sigma}$, $G(a,b)$ denotes a gamma distribution with shape parameter $a$ and rate parameter $b$, and $IG(a,b)$ denotes an inverse gamma distribution with shape parameter $a$ and scale parameter $b$. Additionally, we will use $\pi(\cdot)$ as general notation for a prior probability measure and $\pi(\cdot\mid\boldsymbol{y})$ as general notation for a posterior probability measure.

\subsection{Group Inverse-Gamma Gamma (GIGG) Prior}

Consider a Bayesian sparse linear regression model \begin{equation}
    [\boldsymbol{y} | \boldsymbol{\alpha}, \boldsymbol{\beta}, \sigma^2] \sim N\bigg(\boldsymbol{C}\boldsymbol{\alpha} + \sum_{g=1}^{G}\boldsymbol{X}_g\boldsymbol{\beta}_g,\sigma^2\boldsymbol{I}_{n}\bigg), \hspace{2 mm} \pi(\boldsymbol{\alpha}) \propto 1, \hspace{2 mm} \boldsymbol{\beta} \sim \pi(\boldsymbol{\beta}), \hspace{2 mm} \pi(\sigma^2) \propto \sigma^{-2},
    \label{eqn:linear_model}
\end{equation} where $g = 1,\ldots,G$ indexes the groups, $\boldsymbol{y}$ is an $n \times 1$ vector of centered continuous responses, $\boldsymbol{C}$ is a matrix of adjustment covariates, $\boldsymbol{X}_g$ is an $n \times p_g$ matrix of standardized covariates in the $g$-th group, $\boldsymbol{\beta}_g = (\beta_{g1},\ldots, \beta_{gp_g})^{\top}$ is a $p_g \times 1$ vector of regression coefficients corresponding to the $g$-th group,  $\boldsymbol{\beta} = (\boldsymbol{\beta}^{\top}_1,\ldots,\boldsymbol{\beta}^{\top}_G)^{\top}$ is a $p \times 1$ vector of regression coefficients to employ shrinkage on, and $\boldsymbol{I}_{n}$ is an $n \times n$ identity matrix. We assume the model is sparse in the sense that many of the entries in  $\boldsymbol{\beta}$ are zero. The group inverse-gamma gamma (GIGG) prior is defined as $$[\beta_{gj}|\tau^2,\gamma_g^2,\lambda_{gj}^2] \sim N(0,\tau^2\gamma_g^2\lambda_{gj}^2), \hspace{2 mm} [\gamma_g^2|a_g] \sim G(a_g,1), \hspace{2 mm} [\lambda_{gj}^2|b_g] \sim IG(b_g,1), \hspace{2 mm} [\tau^2,\sigma^2] \sim \pi(\tau^2,\sigma^2),$$ where $j = 1,\ldots,p_g$ indexes the covariates within the $g$-th group. Alternatively, we may also express the prior on $\boldsymbol{\beta}$ as a vector, $[\boldsymbol{\beta}|\tau^2,\boldsymbol{\Gamma},\boldsymbol{\Lambda}] \sim N(0, \tau^2\boldsymbol{\Gamma}\boldsymbol{\Lambda})$, where $\boldsymbol{\Lambda} = \text{diag}(\lambda_{11}^2,...,\lambda_{Gp_G}^2)$ and $\boldsymbol{\Gamma} = \text{diag}(\gamma_1^2,...,\gamma_1^2,\gamma_2^2,...,\gamma_2^2,...,\gamma_G^2,...,\gamma_G^2)$ such that $\gamma_g^2$ is repeated $p_g$ times along the diagonal of $\boldsymbol{\Gamma}$. In the GIGG prior specification, the priors on the group shrinkage parameter, $\gamma_g^2$, and local shrinkage parameter, $\lambda_{gj}^2$, are selected such that the induced prior on the product is a beta prime prior, $\gamma_g^2\lambda_{gj}^2 \sim \beta'(a_g,b_g)$ (see Appendix \ref{app:dist_def} for distributional definitions). Since the group shrinkage parameter is shared by all $p_g$ observations in the $g$-th group, assigning a beta prime prior on the product allows for normal beta prime shrinkage marginally such that the shrinkage is correlated within group. One point that deserves further clarification is the assignment of the gamma and inverse-gamma priors to the group and local parameters, respectively, when either configuration would yield a beta prime prior in the product. The rationale behind this choice is that the inverse-gamma prior is heavier-tailed than the gamma prior, thereby preventing overregularization of large, non-null coefficients due to being grouped with null coefficients. Setting $a_g = b_g = 1/2$ for all $g$ yields a special case of the GIGG prior called the group horseshoe prior, which has correlated horseshoe regularization within group. For a group of size one, the group shrinkage parameter becomes a local shrinkage parameter and we recover the horseshoe prior from the group horseshoe prior.

\subsection{Marginal Prior Properties}

When discussing a proposed shrinkage prior on $\boldsymbol{\beta}$, there are two key features of the marginal prior that need to be investigated. The first is the behavior in a tight neighborhood around zero and the second is the rate at which the prior decays in the extremes. For $\tau^2 = 1$ fixed, \cite{bai2019} showed that the marginal prior $\pi(\beta_{gj} \mid \tau^2,a_g,b_g)$ has a pole at $0$ if and only if $0 < a_g \leq 1/2$, with the pole at zero becoming stronger the closer $a_g$ is to zero. Therefore, one should select $a_g \in (0,1/2]$ for sparse estimation problems to sufficiently shrink null coefficients towards zero. To clarify the tail behavior we need to introduce the notion of a regularly varying function \citep{bingham1989}: A positive, measurable function $f$ is said to be regularly varying at $\infty$ with index $\omega \in \mathbb{R}$ if $\lim_{x \to \infty}f(tx)/f(x) = t^{\omega}$, for all $t > 0$.

\begin{theorem} \label{thm:tail}
Let $\mathcal{B}(a_g,b_g)$ denote the beta function evaluated at $a_g$ and $b_g$ and $\Gamma(b_g+1/2)$ denote the gamma function evaluated at $b_g+1/2$. The tails of the marginal prior probability density function of $\beta_{gj}$ decay at the following rate, $$\lim_{\beta_{gj} \to \infty}\frac{\pi(\beta_{gj} \mid \tau^2,a_g,b_g)}{r(\beta_{gj},\tau^2,a_g,b_g)} = 1, \hspace{2 mm} r(\beta_{gj},\tau^2,a_g,b_g) = \frac{(2\tau^2)^{b_g}\Gamma(b_g+1/2)}{\sqrt{\pi}\mathcal{B}(a_g,b_g)}|\beta_{gj}|^{-(1+2b_g)}\bigg(\frac{\beta_{gj}^2/\tau^2}{1+\beta_{gj}^2/\tau^2}\bigg)^{a_g}.$$ Consequently, the index of regular variation is $\omega = -1-2b_g$.
\end{theorem}

\noindent \textit{Proof.} See Appendix \ref{proof:tail}.

The concept of regular variation has been extensively discussed in the context of Bayesian robustness and noninformative inference \citep{dawid1973, ohagan1979, andrade2006}, with the latter being recently elaborated on in the context of global-local shrinkage priors \citep{bhadra2016}. When the index $\omega < 0$, regular variation essentially states that the tail of the function decays at a polynomial rate and is therefore considered heavy-tailed. Some examples of priors with regularly varying tails include the student's t prior and the horseshoe prior. Conversely, commonly used priors such as the normal prior and the Laplace prior do not have regularly-varying tails. As a consequence of having exponentially decaying tails, Bayesian linear regression with independent normal priors and Bayesian lasso are prone to overregularizing large signals and are not flexible enough to facilitate conflict resolution between discordant likelihood and prior information \citep{andrade2006, polsonscott2011}. Theorem \ref{thm:tail} shows that for any pair of hyperparameters $a_g$ and $b_g$, the marginal GIGG prior has regularly varying tails and, furthermore, that $b_g$ controls the rate at which the tails decay.


\subsection{Sparse Normal Means}

To further elucidate the shrinkage profile of the GIGG prior, we will focus on a special case of the sparse linear regression model called the sparse normal means model ($\boldsymbol{X} = \boldsymbol{I}_n$ and $\boldsymbol{C}$ empty). In the global-local shrinkage prior literature, it is conventional to work with the sparse normal means problem for analytical tractability, even when the ultimate goal is regression \citep{rockova2014, bhattacharya2015}, as the posterior mean has a convenient representation, $E[\beta_{gj} \mid y_{gj},\tau^2,\sigma^2] = (1-E[\kappa_{gj} \mid y_{gj},\tau^2,\sigma^2])y_{gj}$. Here, $\kappa_{gj} = \sigma^2/(\sigma^2+\tau^2\gamma_g^2\lambda_{gj}^2)$ is called a shrinkage factor, because it quantifies how much the posterior mean is shrunk relative to the maximum likelihood estimator $y_{gj}$. Calculating the joint prior distribution for the shrinkage factors in the $g$-th group, $\boldsymbol{\kappa}_g = (\kappa_{g1},...,\kappa_{gp_g})^{\top}$, we have

$\pi\big(\boldsymbol{\kappa}_g \mid \tau^2,\sigma^2,a_g,b_g\big) = $ $$\frac{\Gamma(a_g+p_gb_g)}{\Gamma(a_g)\big(\Gamma(b_g)\big)^{p_g}}\bigg(\frac{\tau^2}{\sigma^2}\bigg)^{p_gb_g}\Bigg(1 + \frac{\tau^2}{\sigma^2}\sum_{j=1}^{p_g}\frac{\kappa_{gj}}{1-\kappa_{gj}}\Bigg)^{-(a_g+p_gb_g)}\Bigg(\prod_{j=1}^{p_g}\kappa_{gj}^{b_g-1}(1-\kappa_{gj})^{-(b_g+1)}\Bigg),$$ where $0 < \kappa_{gj} < 1$ for all $1 \leq j \leq p_g$. Evaluating the prior distribution of $\boldsymbol{\kappa}_g$, we see that the joint density multiplicatively factorizes into ``dependent" and ``independent" parts where the degree to which the within-group shrinkage is correlated is governed by the $\sum_{j=1}^{p_g}\kappa_{gj}/(1-\kappa_{gj})$ term. That is, as $a_g+p_gb_g$ goes to zero, the regularization is highly individualistic, whereas if $a_g+p_gb_g$ moves away from zero, then the shrinkage becomes increasingly more correlated within the $g$-th group.

\begin{figure}[!ht]
    \centering
    \includegraphics[scale=1.0, height = 0.45\textheight, width = 1.0\linewidth]{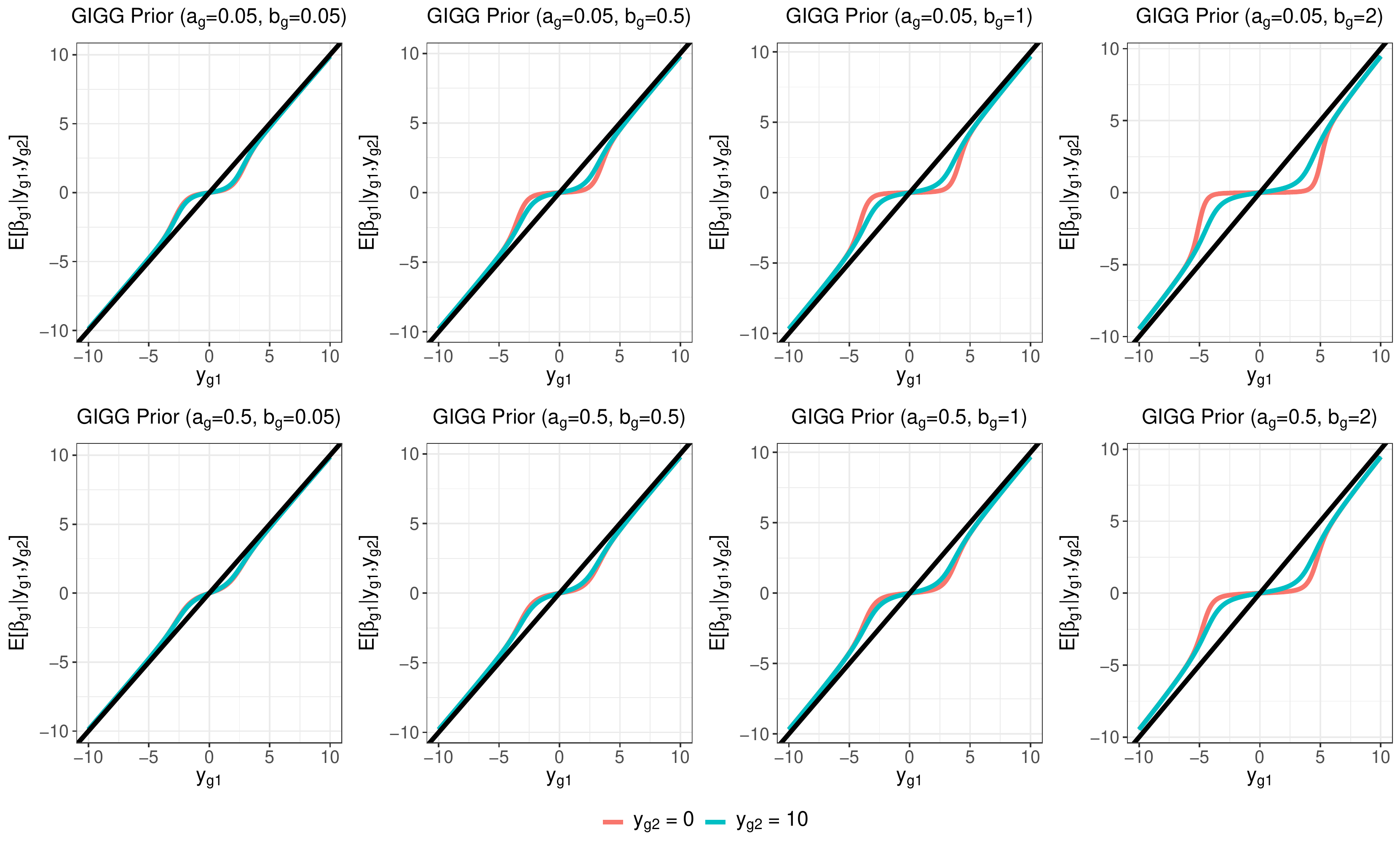}
    \caption{Marginal posterior mean of $\beta_{g1}$ for a group with two observations as $a_g$, $b_g$, $y_{g1}$, and $y_{g2}$ vary. Here, $\tau^2 = 0.2$ and $\sigma^2 = 1$ are fixed.}
    \label{fig:gigg_posterior_mean}
\end{figure}

Figure \ref{fig:gigg_posterior_mean} illustrates the marginal posterior mean of $\beta_{g1}$ for a group of size two as a function of $a_g$, $b_g$, $y_{g1}$, and $y_{g2}$. When $a_g$ and $b_g$ are close to zero then the thresholding effect on the marginal posterior mean of $\beta_{g1}$ hardly depends on the value of $y_{g2}$, indicating highly individualistic shrinkage. This corroborates our intuition from looking at the joint posterior distribution of the shrinkage weights within the same group. The second major observation is that as $b_g$ moves away from zero, the marginal posterior mean of $\beta_{g1}$ becomes increasingly more dependent on the value of $y_{g2}$. In particular, if we look at the case when $a_g = 0.05$ and $b_g = 2$, we see that when $y_{g2} = 0$ the thresholding effect on $\beta_{g1}$ is much stronger when compared to $y_{g2} = 10$. The last major observation is that as $a_g$ moves away from $0$, the thresholding effect becomes weaker. Therefore, $a_g$ effectively controls the overall strength of the shrinkage, whereas $b_g$ generally controls the dependence of the within-group shrinkage.

\section{THEORETICAL PROPERTIES} \label{sec:theory}

\subsection{Linear Regression} \label{subsec:post_consist}

Let $\boldsymbol{X}_n = [\boldsymbol{X}_1,\ldots,\boldsymbol{X}_{G_n}]$ and $\mathcal{H}_n = \{\boldsymbol{a}, \boldsymbol{b}\}$ denote the collection of hyperparameters where $\boldsymbol{a} = \{a_{1},...,a_{G_n}\}$ and $\boldsymbol{b} = \{b_{1},...,b_{G_n}\}$. Here, the subscript $n$ in $G_n$ refers to the fact that the number of groups in the covariate space is growing as a function of the sample size. Furthermore, let $\mathcal{A}_n = \{(g,j) : \beta_{gj}^0 \neq 0\}$ denote the true active set with cardinality $|\mathcal{A}_n|$. Then, Theorem \ref{thm:post_consist} states that the posterior distribution of $\boldsymbol{\beta}_n$ under the GIGG prior is consistent a posteriori for the true $\boldsymbol{\beta}_n^0$. Similarly, we add a subscript $n$ to $\boldsymbol{\beta}_n^0$ and $\boldsymbol{\beta}_n$ to indicate that the number of regression coefficients is growing as function of sample size.

\begin{theorem} \label{thm:post_consist}
    Suppose that $p_n = o(n)$, $L_n = \sup_{(g,j)}|\beta_{gj}^0| < \infty$, where $\beta_{gj}^0$ indicates the true $j$-th regression coefficient in the $g$-th group, $0 < \lim_{n \to \infty}\inf{\mathcal{H}_n} \leq \lim_{n \to \infty}\sup{\mathcal{H}_n} < \infty$, and $|\mathcal{A}_n| = o(n/\log(n))$. Further, suppose that the smallest and largest singular values of $\boldsymbol{X}_n$, denoted by $\theta_{n,min}(\boldsymbol{X}_n)$ and $\theta_{n,max}(\boldsymbol{X}_n)$, satisfy $0 < \liminf_{n \to \infty} \theta_{n,min}(\boldsymbol{X}_n)/\sqrt{n} \leq \limsup_{n \to \infty} \theta_{n,max}(\boldsymbol{X}_n)/\sqrt{n} < \infty$. Then for any $\epsilon > 0$, $$\pi_n(\boldsymbol{\beta}_n : \Vert\boldsymbol{\beta}_n-\boldsymbol{\beta}_n^0\Vert_2 < \epsilon \mid \boldsymbol{y}_n, \mathcal{H}_n, \tau_n^2, \sigma^2) \to 1$$ almost surely as $n \to \infty$ provided that $\tau_n^2 = C/(p_nn^{\rho}\log(n))$ for some $\rho,C \in (0,\infty)$.
\end{theorem}

\noindent\textit{Proof.} See Appendix \ref{proof:post_consist}.

Of note, the only restrictions placed on the values of the hyperparameters in Theorem \ref{thm:post_consist} are that they do not converge to the boundary of the hyperparameter space as $n \to \infty$. 
\begin{remark} Theorem \ref{thm:post_consist} is a generalization of Theorem 5 in \cite{armagan2013} which proved posterior consistency for the NBP prior when $b_g \in (1,\infty)$. Restricting $b_g \in (1,\infty)$ was done to utilize an argument which required the existence of the second moment of $\beta_{gj}$, but does not cover special cases of particular interest such as the horseshoe prior. Therefore, our result extends the existing posterior consistency result from \cite{armagan2013} to a more general collection of hyperparameter values with potential grouping structure.
\end{remark}

Next, we partially extend the posterior concentration theoretical framework for the sparse normal means model, developed in Section 3.2 of \cite{datta2013}, to a low-dimensional linear regression ($p < n$) model with general correlation structure. Going forward, we will drop the subscript $n$ from the notation introduced in the statement of Theorem \ref{thm:post_consist} to clarify that the subsequent theoretical results hold for fixed $p$.

\begin{theorem} \label{thm:concentration_regression}
Fix $\epsilon \in (0,1)$, $p$, and $n$, such that $p < n$. Further, suppose that the smallest and largest singular values of $\boldsymbol{X}^{\top}\boldsymbol{X}$, denoted by $\theta_{min}(\boldsymbol{X}^{\top}\boldsymbol{X})$ and $\theta_{max}(\boldsymbol{X}^{\top}\boldsymbol{X})$, satisfy $0 < \theta_{min}(\boldsymbol{X}^{\top}\boldsymbol{X}) \leq \theta_{max}(\boldsymbol{X}^{\top}\boldsymbol{X}) < \infty$. The full conditional posterior mean corresponding to the GIGG prior is, $$E[\boldsymbol{\beta} \mid \cdot] = \bigg(\boldsymbol{I}_{p}+(\boldsymbol{X}^{\top}\boldsymbol{X})^{-1}\frac{\sigma^2}{\tau^2}\boldsymbol{\Gamma}^{-1}\boldsymbol{\Lambda}^{-1}\bigg)^{-1}\hat{\boldsymbol{\beta}}^{OLS}, \hspace{2 mm} \hat{\boldsymbol{\beta}}^{OLS} = (\boldsymbol{X}^{\top}\boldsymbol{X})^{-1}\boldsymbol{X}^{\top}\boldsymbol{y}.$$ Then the inequality, $$\Big\Vert\hat{\boldsymbol{\beta}}^{OLS}-E[\boldsymbol{\beta} \mid \cdot]\Big\Vert_2 \geq \bigg(\frac{1}{1+\theta_{max}(\boldsymbol{X}^{\top}\boldsymbol{X})\sigma^{-2}\tau^2\max_{(g,j)}\gamma_g^2\lambda_{gj}^2}\bigg)\Big\Vert\hat{\boldsymbol{\beta}}^{OLS}\Big\Vert_2,$$ holds and we have the following results:
\begin{enumerate}[label=\alph*)]
\item $$\pi\bigg(\frac{1}{1+\theta_{max}(\boldsymbol{X}^{\top}\boldsymbol{X})\sigma^{-2}\tau^2\max_{(g,j)}\gamma_g^2\lambda_{gj}^2} \geq \epsilon\ \bigg|\ \boldsymbol{y},\mathcal{H},\tau^2,\sigma^2\bigg) \to 1 \hspace{2 mm} \text{ as } \tau^2 \to 0.$$
\item $$\pi\bigg(\Big\Vert\hat{\boldsymbol{\beta}}^{OLS}-E[\boldsymbol{\beta} \mid \cdot]\Big\Vert_2 \geq \epsilon\Big\Vert\hat{\boldsymbol{\beta}}^{OLS}\Big\Vert_2\ \bigg|\ \boldsymbol{y},\mathcal{H},\tau^2,\sigma^2\bigg) \to 1 \hspace{2 mm} \text{ as } \tau^2 \to 0.$$
\end{enumerate}
\end{theorem}

\noindent\textit{Proof.} See Appendix \ref{proof:concentration_regression}.

Theorem \ref{thm:concentration_regression} states that, irrespective of the correlation structure, $\tau^2 \to 0$ sufficiently shrinks the posterior mean towards zero. The argument used in the proof of Theorem \ref{thm:concentration_regression} can be applied to a litany of other continuous shrinkage priors for which existing posterior concentration results are limited to the sparse normal means model. To supplement these results, we consider the case where we have block diagonal correlation structure, with the blocks defined by the groups, as in Figure \ref{fig:ewas_corrplot}.

\begin{corollary} \label{thm:concentration_regression_b}
Suppose that the covariates in $\boldsymbol{X}$ satisfy $\boldsymbol{X}_g^{\top}\boldsymbol{X}_{g'} = \boldsymbol{0}$ for all $g \neq g'$, where $\boldsymbol{0}$ denotes a $p_g \times p_{g'}$ matrix of zeros. If $\tau^2$, $\sigma^2$, and $a_g \in (0,1)$ are fixed, then there exists a constant $$\epsilon_g(\tau^2,\sigma^2) = \frac{\sigma^2}{\sigma^2+\theta_{max}(\boldsymbol{X}_g^{\top}\boldsymbol{X}_g)\tau^2},$$ such that for all $\delta \in (0, \epsilon_g(\tau^2,\sigma^2))$
$$\pi\bigg(\Big\Vert\hat{\boldsymbol{\beta}}_g^{OLS}-E[\boldsymbol{\beta}_g \mid \cdot]\Big\Vert_2 \geq \delta\Big\Vert\hat{\boldsymbol{\beta}}_g^{OLS}\Big\Vert_2\ \bigg{|}\  \boldsymbol{y},\mathcal{H},\tau^2,\sigma^2\bigg) \to 1$$ as $b_g \to \infty$.
\end{corollary}

\noindent\textit{Proof.} See Appendix \ref{proof:concentration_regression_b}.

The conclusion of Corollary \ref{thm:concentration_regression_b} is that if the hyperparameter $b_g \to \infty$ then there is at least some amount of shrinkage relative to the ordinary least squares estimator in the $g$-th group. If $\tau^2/\sigma^2$ is close to zero, then $\epsilon(\tau^2,\sigma^2) \approx 1$, implying shrinkage of the posterior mean towards zero. Therefore, we can interpret the case when $b_g \to \infty$ and $\tau^2/\sigma^2$ close to zero as shrinkage of the entire $g$-th group towards zero.

\subsection{Sparse Normal Means} \label{subsec:post_concentrate}


Although we would ideally consider additional posterior concentration results within the context of a linear regression model, there is not an analytically tractable analog of componentwise shrinkage factors for a general design matrix without any orthogonality. Therefore, we will proceed by considering posterior concentration results within the sparse normal means framework, to make precise statements regarding componentwise shrinkage, as opposed to shrinkage of the entire $L_2$-norm.

One question that arises is whether the dependence induced between the $\beta_{gj}$'s by $\gamma_g^2$ will overly dominate the individual-level shrinkage. As an example, one can conceptualize a case where a group has only one signal, which is overly shrunk by virtue of being grouped with an overwhelming majority of null means. Theorem \ref{thm:post_concentrate}a states that if the $gl$-th observation is sufficiently large then there will be minimal shrinkage on $y_{gl}$. This guarantees that group shrinkage will not overly dominate individual shrinkage if the observation is large. Conversely, Theorem \ref{thm:post_concentrate}b states that if the global shrinkage parameter converges to zero, then the GIGG prior will sufficiently shrink the $y_{gl}$'s toward zero. Let $\boldsymbol{y}_g = (y_{g1},\ldots,y_{gp_g})^{\top}$.

\begin{theorem} \label{thm:post_concentrate}
Suppose that $p_g \in \{2,3,\ldots\}$..

\begin{enumerate}[label=\alph*)]
    \item Fix $\psi, \delta \in (0,1)$. Then there exists a function $h(p_g,\tau^2,\sigma^2,a_g,b_g,\psi,\delta)$ such that
    $$\pi(\kappa_{gl} > \psi \mid \boldsymbol{y}_g,\tau^2,\sigma^2,a_g,b_g) \leq \exp\bigg({-\frac{\psi(1-\delta)}{2\sigma^2}y_{gl}^2}+{\frac{\psi\delta}{2\sigma^2}\sum_{j \neq l}y_{gj}^2}\bigg)h(p_g,\tau^2,\sigma^2,a_g,b_g,\psi,\delta).$$

    \noindent Consequently, if $|y_{gl}| \to \infty$, then $\pi(\kappa_{gl} \leq \psi \mid \boldsymbol{y}_g,\tau^2,\sigma^2,a_g,b_g) \to 1.$
    \item Fix $\epsilon \in (0,1)$. Then there exists a function $h(p_g,\sigma^2,\boldsymbol{y}_g,a_g,b_g,\epsilon)$ such that,
    $$\pi(\kappa_{gl} < \epsilon \mid \boldsymbol{y}_g,\tau^2,\sigma^2,a_g,b_g) \leq \bigg(\frac{\tau^2}{\sigma^2}\bigg)^{p_g/2+b_g}\Bigg(\min\bigg(1,\frac{\tau^2}{\sigma^2}\bigg)\Bigg)^{-p_g/2}h(p_g,\sigma^2,\boldsymbol{y}_g,a_g,b_g,\epsilon).$$
    
    \noindent Consequently, $\pi(\kappa_{gl} \geq \epsilon \mid \boldsymbol{y}_g,\tau^2,\sigma^2,a_g,b_g) \to 1$ as $\tau^2 \to 0$.
\end{enumerate}
\end{theorem}

\noindent\textit{Proof.} See Appendices \ref{proof:post_concentrate_a} and \ref{proof:post_concentrate_b}.

\vspace{2 mm}

The theoretical statements outlined in Theorem \ref{thm:post_concentrate} were originally discussed for the horseshoe prior \citep{datta2013}, but have also been used in the context of several other continuous shrinkage priors \citep{datta2016, bhadra2017, bai2019}, dynamic trend filtering \citep{kowal2019}, and small area estimation \citep{tang2018sa}. Examining Theorem \ref{thm:post_concentrate}, we first note that neither result restricts the range of values $a_g$ and $b_g$ can take. Therefore, Theorem \ref{thm:post_concentrate} applies to a more general class of hyperparameter values than those considered in \cite{bai2019}. Secondly, the rate at which the upper bound on Theorem \ref{thm:post_concentrate}b converges to zero as $\tau \to 0$ depends on the hyperparameter $b_g$, with larger values of $b_g$ corresponding to a tighter upper bound. To better understand the role of $b_g$ we have the following result.

\begin{corollary} \label{thm:post_group_shrink}
Suppose that $p_g \in \{2,3,\ldots\}$. If $\tau^2$, $\sigma^2$, and $a_g \in (0,1)$ are fixed, then there exists a constant $$\epsilon(\tau^2,\sigma^2,p_g) = \bigg(1+\frac{\tau^2}{\sigma^2}p_g^{p_g}\bigg)^{-1},$$ such that $\pi(\kappa_{gl} < \epsilon(\tau^2,\sigma^2,p_g) \mid \boldsymbol{y}_g, \tau^2,\sigma^2,a_g,b_g) \to 0$ as $b_g \to \infty$.
\end{corollary}

\noindent\textit{Proof.} See Appendix \ref{proof:post_group_shrink}.

\vspace{2 mm}

The conclusion of Corollary \ref{thm:post_group_shrink} is nearly identical to the conclusion of Corollary \ref{thm:concentration_regression_b}, as $\tau^2/\sigma^2$ controls the degree of shrinkage provided as a result of taking $b_g \to \infty$.

\section{COMPUTATION} \label{sec:computation}

\subsection{Gibbs Sampler}

The full conditional updates corresponding to model (\ref{eqn:linear_model}), where $\boldsymbol{\beta}$ is endowed with a GIGG prior, are enumerated in Appendix \ref{app:gibbs_sampler}. Following \cite{polsonscott2011}, we assign a half Cauchy prior scaled by the residual error standard deviation $\tau \mid \sigma \sim C^{+}(0,\sigma)$ and use a prevalent data augmentation trick, $$[\tau^2 \mid \nu] \sim IG(1/2,1/\nu), \hspace{2 mm} [\nu \mid \sigma^2] \sim IG(1/2,1/\sigma^2),$$ to obtain closed form full conditional updates for $\tau^2$ and $\sigma^2$ \citep{makalic2016}. There are two major computational bottlenecks for the proposed algorithm. The first is the full conditional update of $\boldsymbol{\beta}$, $$[\boldsymbol{\beta} \mid \cdot] \sim N\Bigg(\boldsymbol{Q}^{-1}\frac{1}{\sigma^2}\boldsymbol{X}^{\top}\Big(\boldsymbol{y} - \boldsymbol{C}\boldsymbol{\alpha}\Big),\boldsymbol{Q}^{-1}\Bigg), \hspace{2 mm} \boldsymbol{Q} = \frac{1}{\sigma^2}\boldsymbol{X}^{\top}\boldsymbol{X}+\frac{1}{\tau^2}\boldsymbol{\Gamma}^{-1}\boldsymbol{\Lambda}^{-1}.$$ The second occurs when there are a multitude of group and local parameters that need to be drawn at each iteration of the Gibbs sampler, which is often the case in ``large $p$" scenarios. Rather than na\"ively sampling from the full conditional distributions there are several strategies to achieve faster posterior computation with both of these computational challenges in mind:

\begin{itemize}
    \item Draw $\boldsymbol{v} \sim N\big(\sigma^{-2}\boldsymbol{X}^{\top}(\boldsymbol{y} - \boldsymbol{C}\boldsymbol{\alpha}),\boldsymbol{Q}\big)$, and then solve $\boldsymbol{Q}\boldsymbol{\beta} = \boldsymbol{v}$, rather than explicitly calculating $\boldsymbol{Q}^{-1}$.
    \item For ``small $n$, large $p$" problems, the Woodbury identity can be utilized so that the full conditional update of $\boldsymbol{\beta}$ scales linearly in $p$ \citep{bhattacharya2016}.
    \item If $n$ and $p$ are both large, say an order of magnitude of 10,000 each, there are several recently developed approximation approaches, the former of which exploits the ability of the horseshoe prior to shrink $\tau^2\lambda_{gj}^2$ close to zero \citep{johndrow2020} while the latter uses a conjugate gradient algorithm to find an approximate solution to $\boldsymbol{Q}\boldsymbol{\beta} = \boldsymbol{v}$ \citep{nishimura2020}.
    \item Parallelization can be used within the Gibbs sampler to simultaneously update the shrinkage parameters corresponding to each group \citep{terenin2019}.
\end{itemize}

\subsection{Hyperparameter Selection}

If the modeler wants to remain relatively agnostic to the choice of hyperparameters, one can use Marginal Maximum Likelihood Estimation (MMLE) \citep{casella2001}, an empirical-Bayes approach executed iteratively within the Gibbs sampler. The $(l+1)$th update is
$$a_g^{(l+1)} = \psi_0^{-1}\bigg(E_{a_g^{(l)}}\big[\log(\gamma_g^2) \mid \boldsymbol{y}\big]\bigg), \hspace{2 mm} b_g^{(l+1)} = \psi_0^{-1}\bigg(-\frac{1}{p_g}\sum_{j=1}^{p_g}E_{b_g^{(l)}}\big[\log(\lambda_{gj}^2) \mid \boldsymbol{y}\big]\bigg),$$ where $\psi_0(\cdot)$ is the digamma function and the expectation terms can be estimated through standard Monte Carlo methods. The iterative procedure terminates when $\sum_{g=1}^{G}\big(a_g^{(l+1)} - a_g^{(l)}\big)^2 + \sum_{g=1}^{G}\big(b_g^{(l+1)} - b_g^{(l)}\big)^2$ is less than some prespecified error tolerance. However, in our experience it is preferred to fix $a_g = 1/n$ for all $g$ and use MMLE to estimate the $b_g$ hyperparameters. The first reason is that $a_g$ controls the strength of the thresholding effect and choosing $a_g$ close to zero guarantees strong shrinkage of null coefficients towards zero. The second reason is that only estimating one hyperparameter per group is more feasible than estimating two hyperparameters per group, particularly when the number of groups is large. Since $b_g$ primarily controls how the correlated the shrinkage is within-group it is more important to focus estimation on the $b_g$ hyperparameters. We do recognize that setting $a_g = 1/n$ violates a condition in Theorem \ref{thm:post_consist} where the infimum of the set of hyperparameters cannot converge to zero as $n \to \infty$. However, for practical purposes, this approach provides an automatic way to set $a_g$ while also yielding similar results to $a_g$ close to zero and fixed as a function of the sample size, such as $a_g = 1/100$.

Although MMLE is useful for problems where the number of groups, $G$, is small relative to the sample size, the estimates for the $a_g$'s and $b_g$'s will become increasingly variable in high-dimensional settings where the number of groups is large. There may also be low-dimensional settings where the user wants to incorporate explicit prior knowledge about the nature of the within-group signal density. In such cases, it may be preferred to fix hyperparameter values in accordance with subject matter expertise. As with the modified MMLE approach, we recommend setting $a_g = 1/n$ for all $g$. To fix $b_g$ we recommend a useful heuristic whereby local, group, and global shrinkage parameters are simulated from the GIGG prior. Using the simulated shrinkage parameters, shrinkage factors can be constructed and the correlation between shrinkage factors within the same group can be empirically calculated. Selecting the hyperparameter $b_g$ is then equivalent to selecting how correlated the shrinkage is within-group, a more easily understandable concept. Implementations of GIGG regression with fixed hyperparameters and hyperparameters estimated via MMLE are available on \href{https://github.com/umich-cphds/gigg}{Github}. 

\section{SIMULATIONS} \label{sec:simulations}

\subsection{Generative Model}

The data generative mechanism is linear regression model (\ref{eqn:linear_model}), where $\boldsymbol{C}$ includes the intercept term and five adjustment covariates drawn from independent standard normal distributions, $\boldsymbol{\alpha} = (0, 1, 1, 1, 1, 1)^{\top}$, and $\boldsymbol{X}$ is drawn from a multivariate normal distribution with mean $\boldsymbol{0}$ and covariance matrix $\boldsymbol{\Sigma}_{\boldsymbol{X}}$. $\boldsymbol{\Sigma}_{\boldsymbol{X}}$ is determined such that the features have unit variance and block-diagonal exchangeable correlation structure. For all simulation settings, $n = 500$ and $p = 50$ such that the 50 covariates are evenly divided into five groups. Pairwise correlations within each group are $\rho = 0.8$ for the high correlation simulation settings or $\rho = 0.6$ for the medium correlation simulation settings. For all simulation settings, the pairwise correlations across groups are $0.2$ and the residual error variance, $\sigma^2$, is fixed such that $\boldsymbol{\beta}^{\top}\boldsymbol{\Sigma}_{\boldsymbol{X}}\boldsymbol{\beta}/(\boldsymbol{\beta}^{\top}\boldsymbol{\Sigma}_{\boldsymbol{X}}\boldsymbol{\beta}+\sigma^2) = 0.7$.

For the fixed regression coefficient simulation settings, we consider two realizations of $\boldsymbol{\beta}$, which will be qualitatively referred to as the concentrated signal setting and the distributed signal setting. In the concentrated signal setting, there is only one true signal in each of the five groups with varying magnitudes: $\beta_{11} = 0.5$, $\beta_{21} = 1$, $\beta_{31} = 1.5$, $\beta_{41} = 2$, and $\beta_{51} = 2$. Rather than having within-group sparsity, the distributed signal setting assumes that the signal is shared across all members of the first group: $\beta_{1j} = 0.5$ for all $j \in \{1,...,5\}$ and $\beta_{1j} = 1$ for all $j \in \{6,...,10\}$. The purpose of the fixed coefficient simulation settings is to ascertain which methods perform well when the within-group signal is sparse or dense.

Beyond the fixed regression coefficient simulation settings, we also consider a {\it random coefficient} simulation in the high correlation setting, where for each simulation iteration a random regression coefficient vector is generated. To construct a regression coefficient vector, we start by randomly selecting either a concentrated or distributed signal for the first group with even probability to guarantee that each simulation iteration will have at least one true signal. The concentrated and distributed signal magnitudes are selected such that the contribution to $\boldsymbol{\beta}^{\top}\boldsymbol{\Sigma}_{\boldsymbol{X}}\boldsymbol{\beta}$ is equal, namely the distributed signal is $\beta_{gj} = 0.25$ for $j = 1,...,10$ and the concentrated signal is $\beta_{g1} = 5.125$ and $\beta_{gj} = 0$ for $j = 2,...,10$. For the other four groups, we randomly select a concentrated signal with probability 0.2, a distributed signal with probability 0.2, and no signal with probability 0.6. The goal of the random coefficient simulation setting is to show that, averaged across many combinations of regression coefficient vectors comprised of sparse within-group signals, dense within-group signals, and inactive groups, GIGG regression with MMLE results in the lowest mean-squared error.

\subsection{Competing Methods and Evaluation Metrics}

Estimation properties will be evaluated based on empirical mean-squared error (MSE), stratified by null and non-null coefficients, across 5000 replicates. In the random coefficient simulations, calculating the MSE corresponds to an integrated mean-squared error (IMSE) metric averaged across the generative distribution of the regression coefficient vectors. For the fixed coefficient simulations we will consider several special cases of the GIGG prior with fixed hyperparameters, namely all possible combinations of $a_g \in \{1/n,1/2\}$ and $b_g \in \{1/n,1/2,1\}$. That way, we can check whether the intuition gleaned from Figure \ref{fig:gigg_posterior_mean} empirically translates to the regression setting. We will also consider the GIGG prior when the hyperparameters $a_g = 1/n$ are fixed and $b_g$ are estimated via MMLE. 

The list of competing methods include Ordinary Least Squares (OLS), Horseshoe regression, Group horseshoe regression \citep{xu2016}, Spike-and-Slab Lasso \citep{rockova2018}, Bayesian Group Lasso with Spike-and-Slab Priors (BGL-SS) \citep{xu2015}, and Bayesian Sparse Group Selection with Spike-and-Slab Priors (BSGS-SS) \citep{xu2015}. To avoid confusion with the group horseshoe prior proposed in this paper, we will refer to the group horseshoe prior from \cite{xu2016} as the group half Cauchy prior throughout the rest of the simulation section. Most methods requiring Markov chain Monte Carlo (MCMC) sampling have 10000 burn-in draws, followed by 10000 posterior draws with no thinning. The only exceptions are BGL-SS and BSGS-SS which have 1000 burn-in draws and 2000 posterior draws without any thinning, due to the relatively slower posterior sampling algorithms.

\subsection{Simulation Results}

\begin{table}[h]
\begin{center}
\scalebox{1}{
     \begin{tabular}{lcccc}
        \hline\hline
        $\pmb{\rho = 0.8}$ & \multicolumn{2}{c}{\textbf{Concentrated}} & \multicolumn{2}{c}{\textbf{Distributed}} \\
        \cline{2-5}
        \textbf{Method} & Null & Non-Null & Null & Non-Null\\
        \hline
        Ordinary Least Squares & 3.74 & 0.41 & 8.09 & 2.03 \\
        Horseshoe & 0.51 & 0.41 & 0.85 & 2.14 \\
        GIGG ($a_g = 1/n, b_g = 1/n$) & \textbf{0.11} & \textbf{0.30} & 0.03 & 3.59 \\
        GIGG ($a_g = 1/2, b_g = 1/n$) & \textbf{0.11} & \textbf{0.30} & 0.04 & 3.56 \\
        GIGG ($a_g = 1/n, b_g = 1/2$) & 0.29 & 0.39 & \textbf{0.03} & \textbf{1.57} \\
        *GIGG ($a_g = 1/2, b_g = 1/2$) & 0.33 & 0.40 & 0.24 & 1.70 \\
        GIGG ($a_g = 1/n, b_g = 1$) & 0.53 & 0.49 & \textbf{0.03} & \textbf{1.43} \\
        GIGG ($a_g = 1/2, b_g = 1$) & 0.58 & 0.49 & 0.26 & 1.43 \\
        GIGG (MMLE) & \textbf{0.20} & \textbf{0.34} & \textbf{0.04} & \textbf{1.42} \\
        Group Half Cauchy & 0.30 & 0.39 & 0.08 & 1.64 \\
        Spike-and-Slab Lasso & \textbf{0.15} & \textbf{0.33} & 0.21 & 4.27 \\
        BGL-SS & 2.01 & 0.80 & \textbf{0.04} & \textbf{1.31} \\
        BSGS-SS & 0.23 & 0.42 & 0.04 & 1.84 \\
        \hline
    \end{tabular}
    }
    \caption{Mean-squared errors (MSE) for the fixed regression coefficient simulation settings ($n = 500, p = 50$) with high pairwise correlations ($\rho = 0.8$). Bolded cells indicate the four methods with the lowest overall MSE. *GIGG regression with $a_g = 1/2$ and $b_g = 1/2$ is equivalent to group horseshoe regression.}
    \label{tbl:low_dim_high_corr}
\end{center}
\end{table}

Table \ref{tbl:low_dim_high_corr} presents the MSE for the high correlation simulation settings and Table \ref{tbl:low_dim_medium_corr} lists the MSE for the medium correlation simulation settings. Because the results for the high correlation and medium correlation settings are similar we will only focus our discussion around the high correlation simulation settings. The first noteworthy observation is that group horseshoe regression has a uniformly lower MSE than both OLS and horseshoe regression for both null and non-null estimation, although the discrepancy between horseshoe and OLS is much larger than the difference between group horseshoe and horseshoe, particularly for the null coefficients. For GIGG regression with fixed hyperparameters, the top performer is GIGG regression with $b_g = 1/n$ when the signal is concentrated within-group (Null MSE = 0.11, Non-Null MSE = 0.30) and $a_g = 1/n, b_g = 1$ when the signal is distributed within-group (Null MSE = 0.03, Non-Null MSE = 1.43), exactly as Figure \ref{fig:gigg_posterior_mean} suggests. However, if the user sets $b_g = 1$ when the signal is concentrated (Null MSE = 0.53, Non-Null MSE = 0.49) or $b_g = 1/n$ when the signal is distributed (Null MSE = 0.03, Non-Null MSE = 3.59), then the ``incorrect" prior information results in notably worse MSE compared to the ``correct" prior information. That being said, $b_g = 1/2$ appears to be a middle ground where the performance for both concentrated and distributed simulation settings is generally good.

Examining the performance of the competing methods, we note that Spike-and-Slab Lasso does very well for the concentrated signal setting (Non-Null MSE = 0.33), but struggles when the signal is distributed (Non-Null MSE = 4.27). Conversely, BGL-SS does poorly when the signal is concentrated (Non-Null MSE = 0.80), but has good performance when the signal is distributed (Non-Null MSE = 1.31). Group half Cauchy regression and BSGS-SS have relatively low MSE across all three simulation settings, however, GIGG with MMLE outperforms both methods in the concentrated and distributed simulation settings.

\begin{table}[h]
\begin{center}
\scalebox{1}{
     \begin{tabular}{lcccc}
        \hline\hline
        $\pmb{\rho = 0.6}$ & \multicolumn{2}{c}{\textbf{Concentrated}} & \multicolumn{2}{c}{\textbf{Distributed}} \\
        \cline{2-5}
        \textbf{Method} & Null & Non-Null & Null & Non-Null\\
        \hline
        Ordinary Least Squares & 1.88 & 0.21 & 3.20 & 0.79 \\
        Horseshoe & 0.29 & 0.21 & 0.52 & 0.94 \\
        GIGG ($a_g = 1/n, b_g = 1/n$) & \textbf{0.05} & \textbf{0.19} & 0.04 & 1.52 \\
        GIGG ($a_g = 1/2, b_g = 1/n$) & \textbf{0.05} & \textbf{0.20} & 0.04 & 1.50 \\
        GIGG ($a_g = 1/n, b_g = 1/2$) & 0.15 & 0.22 & \textbf{0.03} & \textbf{0.69} \\
        *GIGG ($a_g = 1/2, b_g = 1/2$) & 0.18 & 0.21 & 0.16 & 0.73 \\
        GIGG ($a_g = 1/n, b_g = 1$) & 0.29 & 0.26 & \textbf{0.02} & \textbf{0.65} \\
        GIGG ($a_g = 1/2, b_g = 1$) & 0.33 & 0.25 & 0.16 & 0.66 \\
        GIGG (MMLE) & \textbf{0.10} & \textbf{0.20} & \textbf{0.03} & \textbf{0.63} \\
        Group Half Cauchy & 0.17 & 0.21 & 0.07 & 0.71 \\
        Spike-and-Slab Lasso & \textbf{0.03} & \textbf{0.25} & 0.01 & 2.18 \\
        BGL-SS & 1.25 & 0.42 & \textbf{0.01} & \textbf{0.61} \\
        BSGS-SS & 0.10 & 0.22 & 0.01 & 0.81 \\
        \hline
    \end{tabular}
    }
    \caption{Mean-squared errors (MSE) for the fixed regression coefficient simulation settings ($n = 500, p = 50$) with medium pairwise correlations ($\rho = 0.6$). Bolded cells indicate the three methods with the lowest overall MSE. *GIGG regression with $a_g = 1/2$ and $b_g = 1/2$ is equivalent to group horseshoe regression.}
    \label{tbl:low_dim_medium_corr}
\end{center}
\end{table}

\begin{table}[h]
\begin{center}
\scalebox{1}{
     \begin{tabular}{lcc}
        \hline\hline
        \textbf{Method} & Null & Non-Null \\
        \hline
        Ordinary Least Squares & 8.84 & 3.38 \\
        Horseshoe & 0.70 & 1.18 \\
        Group Horseshoe & \textbf{0.39} & \textbf{1.13} \\
        Group Half Cauchy & \textbf{0.36} & \textbf{1.14} \\
        GIGG (MMLE) & \textbf{0.19} & \textbf{1.17} \\
        Spike-and-Slab Lasso & 0.16 & 3.65 \\
        BGL-SS & 2.84 & 2.44 \\
        BSGS-SS & 0.36 & 1.45 \\
        \hline
    \end{tabular}
    }
    \caption{Integrated mean-squared errors (IMSE) for the random regression coefficient simulation setting ($n = 500, p = 50$) with high pairwise correlations ($\rho = 0.8$). Bolded cells indicate the three methods with the lowest overall IMSE.}
    \label{tbl:high_dim}
\end{center}
\end{table}

As with the fixed regression coefficient simulations, group horseshoe (Null IMSE = 0.39) and group half Cauchy regression (Null ISME = 0.36) lead to a substantial improvement in IMSE compared to horseshoe regression (see Table \ref{tbl:high_dim}). However, we also observe that the additional flexibility of GIGG regression to self-adapt to different types of within-group signal distributions results in noticeable improvements in IMSE for the null coefficients (Null IMSE = 0.19). Spike-and-Slab Lasso and BGL-SS struggle in the random coefficient simulation scenario because they are designed to work well only when the signal is concentrated or distributed, respectively, leading to poor average performance.

\section{DATA EXAMPLE} \label{sec:ewas}

The National Health and Nutrition Examination Survey (NHANES) is a collection of studies conducted by the National Center for Health Statistics with the overarching goal of evaluating the health and nutritional status of the United States' populace. Data collection consists of a written survey and physical examination which records demographic, socioeconomic, dietary, and health-related information, including physiological measurements and laboratory tests. We will specifically apply GIGG regression to a subset of 990 adults from the 2003-2004 NHANES cycle with 35 measured contaminants across five exposure classes: metals, phthalates, organochlorine pesticides, polybrominated diphenyl ethers (PBDEs), and polycyclic aromatic hydrocarbons (PAHs). Figure \ref{fig:ewas_corrplot} illustrates the block diagonal correlation structure of these exposures, where areas of high correlation are mostly contained within exposure class. Gamma glutamyl transferase (GGT), an enzymatic marker of liver functionality, will be the outcome of interest. GGT and all environmental exposures were log-transformed to remove right skewness and then subsequently standardized. The final model was adjusted for age (quartiles), sex, body mass index (quartiles), poverty-to-income ratio (quartiles), ethnicity, and urinary creatinine (quartiles).

\begin{figure}[!ht]
    \centering
    \includegraphics[scale=1.0, height = 0.75\textheight, width = 1.0\linewidth]{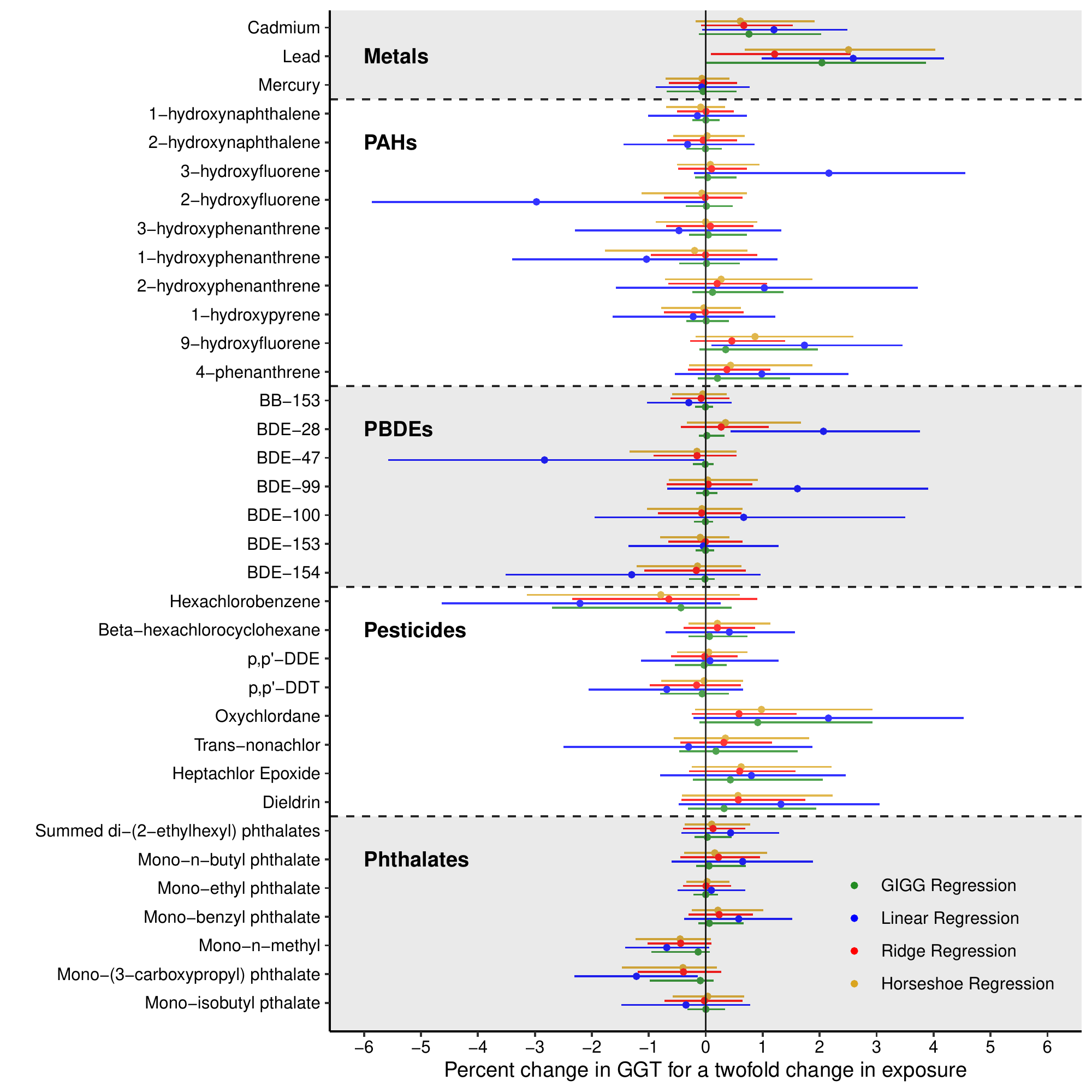}
    \caption{Estimated associations between environmental toxicants (metals, phthalates, pesticides, PBDEs, and PAHs) and gamma glutamyl transferase (GGT) from NHANES 2003-2004 ($n = 990$).}
    \label{fig:ewas_forest_plot_glucose}
\end{figure}

Figure \ref{fig:ewas_forest_plot_glucose} presents the estimated percent change in GGT corresponding to a twofold change in each environmental exposure and their associated 95\% credible intervals for methods commonly used in multipollutant modeling (for a focused comparison of the various group shrinkage methods from the simulation study see Supplementary Figure \ref{fig:ewas_forest_plot_glucose_supp}). Bayesian linear regression with noninformative priors and ridge regression were implemented in R Stan with four chains with no thinning, each with 1000 burn-in draws and 1000 posterior draws. Horseshoe regression and GIGG regression used 10000 burn-in samples, followed by 10000 posterior draws with a thinning interval of five. Convergence of the MCMC chains was evaluated using Gelman-Rubin's potential scale reduction factor (PSRF) \cite{gelman1992}. Bayesian linear regression with noninformative priors, ridge regression, and horseshoe regression had a PSRF of $1.00$ for all regression coefficients and GIGG regression had PSRF values ranging between $1.00 - 1.01$, indicating that all MCMC chains converged. Bayesian linear regression with noninformative priors results in wide credible intervals and highly variable point estimates compared to ridge regression, horseshoe regression, and GIGG regression, particularly in the exposure classes with high pairwise correlations (PAHs and PBDEs). Ridge regularization helps achieve substantial variance reduction compared to linear regression with noninformative priors, but at the cost of overregularizing the apparent signals in the metals exposure class. Namely, ridge regression estimates that a twofold change in lead exposure is associated with 1.21\% higher (95\% CI: 0.09, 2.54) GGT adjusted for age, sex, body mass index, poverty-to-income ratio, ethnicity, and urinary creatinine, while horseshoe regression estimates 2.51\% higher GGT (95\% CI: 0.69, 4.03) and GIGG regression estimates 2.04\% higher GGT (95\% CI: 0.01, 3.87). Although the conclusions between horseshoe regression and GIGG regression are concordant, the difference between the methods can be seen by evaluating the length of credible intervals. The median credible interval length for GIGG regression is 50\% shorter for the PAHs, 79.2\% shorter for the PBDEs, 11.4\% shorter for the pesticides, and 38.9\% shorter for the phthalates compared to horseshoe regression. Note that the exposure classes with markedly narrower credible intervals, PAHs and PBDEs, have high pairwise correlations and common estimated effect sizes. However, the metals exposure class, which has weak pairwise correlations and heterogeneous estimated effect sizes, results in a median credible interval length of 0.31 for GIGG regression and 0.30 for horseshoe regression. The example indicates that by leveraging grouping information GIGG regression has appreciable efficiency gains for groups with multicollinearity issues and homogeneous effect sizes, but does not provide an improvement for groups with weak correlations and heterogeneous effect sizes. Additionally, from a computational perspective, GIGG regression generated a median effective sample size of 590.4 per second, compared to a median effective sample size of 78.9 per second for horseshoe regression from the \textit{horseshoe} package in R.

\section{DISCUSSION} \label{sec:discussion}

The principal methodological contribution of this paper is to construct a continuous shrinkage prior that improves regression coefficient estimation in the presence of grouped covariates. GIGG regression flexibly controls the relative contributions of individual and group shrinkage to improve regression coefficient estimation, resulting in a relative IMSE reduction of 72.8\% corresponding to the null coefficients and 7.5 times more efficient computation compared to the primary horseshoe regression implementation in R. One of the main limitations of GIGG regression is that covariate groupings must be explicitly specified and covariate groupings may not overlap. Additionally, although the GIGG prior can be imposed on regression coefficients in Bayesian generalized linear models, a theoretical evaluation of the shrinkage properties for non-normal outcome data would be necessary to determine if the GIGG prior is appropriate for such models. We are currently working on an R package to implement GIGG regression that, upon completion, will be added to the Comprehensive R Archive Network. For a preliminary version of the R package, visit \href{https://github.com/umich-cphds/gigg}{Github}.

The analysis of multiple pollutant data and chemical mixtures is a key thrust of the National Institute of Environmental Health Sciences, and the GIGG prior provides a useful framework for achieving variance reduction in the presence of group-correlated exposures, characterizing uncertainties in point estimates, and constructing policy relevant metrics, like summary risk scores, in a principled way. However, the generality of the GIGG prior coupled with the relative ease of computation means that, despite its motivation coming from environmental epidemiology, the GIGG prior is applicable to many other areas. For example, in neuroimaging studies, scalar-on-image regression~\citep{kang2018scalar} has been widely used to study the association between brain activity and clinical outcomes of interests. The whole brain can be partitioned into a set of exclusive regions according to brain functions and anatomical structures. Within the same region, the brain imaging biomarkers tend to be more correlated and have similar effects on the outcome variable. The GIGG prior can be extended for scalar-on-image regression and it has a great potential to improve estimating the effects of imaging biomarkers by incorporating brain region information.

In this paper, our focus was sparse estimation, but it is also natural to inquire about uncertainty quantification and variable selection. Based on our simulations, the conclusions of \cite{vanderpas2017} are relevant for the GIGG prior when $0 < a_g \leq 1/2$, but a comprehensive study needs to be carried out. There is no consensus way of defining variable selection for continuous shrinkage priors, however there are several approaches to determine a final active set, including two-means clustering \citep{bhattacharya2015}, credible intervals covering zero \citep{vanderpas2017}, thresholding shrinkage factors \citep{tang2018}, decoupling shrinkage and selection (DSS) \citep{hahn2015}, and penalized credible regions \citep{zhang2018}. For horseshoe-style shrinkage, variable selection defined through credible intervals covering zero is conservative, but works well if one wants to limit the number of false discoveries. The two-means clustering heuristic does not necessarily result in consistent variable selection and the shrinkage factor thresholding approach is restricted to applications where $p < n$. The penalized credible region approach searches for the sparsest model that falls within the $100\times(1-\alpha)$\% joint elliptical credible region, while DSS constructs an adaptive lasso-style objective function with the goal of sparsifying the posterior mean such that most of the predictive variability is still explained. Since the DSS construction is framed from a prediction perspective, this approach may not be ideal for regression coefficient estimation problems in the presence of correlated features. Another crucial point to make is that if one is interested in selection, the posterior mode estimator for the horseshoe prior will result in exact zero estimates, and an approximate algorithm for calculating the posterior mode was developed in \cite{bhadra2019} using the horseshoe-like prior. Therefore, one could conceptualize an extension of the expectation-maximization algorithm developed by \cite{bhadra2019} using a ``GIGG-like" prior. Further work is needed to juxtapose the behavior of all of these different methods for selection and develop novel algorithms for calculating the posterior mode.

\section*{Acknowledgements}

Dr. Datta acknowledges support from the National Science Foundation (DMS-2015460). Dr. Kang acknowledges support from the National Institutes of Health (R01 DA048993; R01 GM124061; R01 MH105561). Dr. Mukherjee acknowledges support from the National Science Foundation (DMS-1712933) and the National Institutes of Health (R01 HG008773-01).

\newpage

\bibliographystyle{apalike}

\bibliography{JASA-template}

\newpage

\section{APPENDICES}

\subsection{Distributions used in Manuscript} \label{app:dist_def}

\noindent Beta Prime Distribution: $$X \sim \beta'(a,b) \implies f_{X}(x) = \frac{\Gamma(a+b)}{\Gamma(a)\Gamma(b)}x^{a-1}(1+x)^{-a-b}, \hspace{2 mm} x > 0.$$

\noindent Gamma Distribution: $$X \sim G(a,b) \implies f_{X}(x) = \frac{b^a}{\Gamma(a)}x^{a-1}\exp(-bx), \hspace{2 mm} x > 0.$$

\noindent Generalized Inverse Gaussian Distribution: $$X \sim GIG(\lambda,\psi,\chi) \implies f_{X}(x) = \frac{(\psi/\chi)^{\lambda/2}}{2K_{\lambda}(\sqrt{\psi\chi})}x^{\lambda-1}\exp\bigg(-\frac{1}{2}\bigg(\frac{\chi}{x}+\psi{x}\bigg)\bigg), \hspace{2 mm} x > 0,$$

\noindent where $K_{\lambda}(\cdot)$ is the modified Bessel function of the third kind with index $\lambda$ \citep{hormann2014}.

\noindent Half Cauchy Distribution: $$X \sim C^{+}(0,\sigma) \implies f_{X}(x) = \frac{2}{\pi\sigma(1+x^2/\sigma^2)}, \hspace{2 mm} x > 0.$$

\noindent Inverse Gamma Distribution: $$X \sim IG(a,b) \implies f_{X}(x) = \frac{b^a}{\Gamma(a)}x^{-a-1}\exp\bigg(-\frac{b}{x}\bigg), \hspace{2 mm} x > 0.$$

\newpage

\subsection{Proof of Theorem \ref{thm:tail}} \label{proof:tail}

For shorthand, we will use: $$r(x) \sim s(x) := \lim_{x \to \infty} \frac{r(x)}{s(x)} = 1.$$ Define $$L(u) = \frac{1}{(\tau^2)^{\lambda}\mathcal{B}(a_g,b_g)}\bigg(\frac{u/\tau^2}{1+u/\tau^2}\bigg)^{a_g}, \hspace{2 mm} \mathcal{B}(a_g,b_g) = \frac{\Gamma(a_g)\Gamma(b_g)}{\Gamma(a_g+b_g)},$$ which is slowly-varying function, i.e., $$\lim_{u \to \infty} \frac{L(tu)}{L(u)} = 1,$$ for all $t > 0$. Moreover, let $$\pi(\beta_{gj} \mid \tau^2,a_g,b_g) = \int_{0}^{\infty}(2{\pi}u)^{-1/2}\exp\bigg(-\frac{\beta_{gj}^2}{2u}\bigg)f(u \mid \tau^2,a_g,b_g)du$$ denote the normal variance mixture probability density function and let $f(u\mid\tau^2,a_g,b_g)$ denote the scaled $\beta'(a_g,b_g)$ mixing density function with fixed scale parameter $\tau^2$. Then $$\lim_{u \to \infty} \frac{f(u\mid\tau^2,a_g,b_g)}{\exp(-\psi_+u)u^{\lambda-1}L(u)}$$ $$= \lim_{u \to \infty} \frac{(\tau^2\mathcal{B}(a_g,b_g))^{-1}(u/\tau^2)^{a_g-1}(1+u/\tau^2)^{-(a_g+b_g)}}{\exp(-\psi_+u)u^{\lambda-1}L(u)} = \lim_{u \to \infty} \frac{(u/\tau^2)^{-1}(1+u/\tau^2)^{-b_g}}{\exp(-\psi_+u)(u/\tau^2)^{\lambda-1}}$$ $$= \lim_{u \to \infty} \exp(\psi_+u)(u/\tau^2)^{-\lambda}(1+u/\tau^2)^{-b_g},$$ where $\psi_+ = \sup\{w \in \mathbb{R} : \phi(w) < \infty\}$ and $$\phi(w) = \frac{1}{\mathcal{B}(a_g,b_g)}\int_{0}^{\infty}\exp(wu)\frac{1}{\tau^2}\bigg(\frac{u}{\tau^2}\bigg)^{a_g-1}\bigg(1+\frac{u}{\tau^2}\bigg)^{-(a_g+b_g)}du.$$ Note that $\psi_+ = 0$. Fix $\lambda = -b_g$. Then, $$\lim_{u \to \infty} \exp(\psi_+u)(u/\tau^2)^{-\lambda}(1+u/\tau^2)^{-b_g} = \lim_{u \to \infty} \bigg(\frac{u/\tau^2}{1+u/\tau^2}\bigg)^{b_g} = 1.$$

\noindent By Theorem 6.1 in \cite{barndorff1982} we conclude that $$\pi(\beta_{gj}\mid\tau^2,a_g,b_g) \sim (2\pi)^{-1/2}2^{b_g+1/2}\Gamma(b_g+1/2)|\beta_{gj}|^{-(1+2b_g)}L(\beta_{gj}^2).$$

\noindent To get the index of regular variation, we note the following straightforward lemma:

\begin{lemma} \label{lemma:rates}
Suppose that $r$ and $s$ are two positive, measurable functions such that $r(x) \sim s(x)$ and $s$ is regularly varying with index $\omega \in \mathbb{R}$. Then $r$ is regularly varying with index $\omega$.
\end{lemma}

\noindent \textit{Proof.} $$\lim_{x \to \infty} \frac{r(tx)}{r(x)} = \lim_{x \to \infty} \frac{r(tx)}{r(x)}\frac{s(tx)}{s(tx)}\frac{s(x)}{s(x)} = \lim_{x \to \infty} \bigg(\frac{r(tx)/s(tx)}{r(x)/s(x)}\bigg)\frac{s(tx)}{s(x)} = \lim_{x \to \infty}\frac{s(tx)}{s(x)} = t^{\omega}.$$

\noindent Despite its simplicity, Lemma \ref{lemma:rates} is of great practical use, particularly if the function whose tail behavior we are interested in does not have a closed form. When working with global-local mixture priors we often do not have closed form marginal prior distributions for $\boldsymbol{\beta}$ and it is usually easier to construct and work with a closed form function that has asymptotically equivalent tail behavior. Since the index of regular variation of $$(2\pi)^{-1/2}2^{b_g+1/2}\Gamma(b_g+1/2)|\beta_{gj}|^{-(1+2b_g)}L(\beta_{gj}^2)$$ is $\omega = -1-2b_g$, then by Lemma \ref{lemma:rates} the index of regular variation of $\pi(\beta_{gj}\mid\tau^2,a_g,b_g)$ is also $\omega = -1-2b_g$.

\newpage

\subsection{Proof of Theorem \ref{thm:post_consist}} \label{proof:post_consist}

Let $\mathcal{G}_n = \{\gamma_{1}^2,...,\gamma_{G_n}^2\}$ denote the collection of group shrinkage parameters and let $p_{g}$ indicate the number of covariates in the $g$-th group. Define the following sets: $\mathcal{A}_{g} = \{j : \beta_{gj}^0 \neq 0\}$, $\mathcal{A}_{g}^c = \{j : \beta_{gj}^0 = 0\}$, $\mathcal{A}_n = \{(g,j) : \beta_{gj}^0 \neq 0\}$, $\mathcal{A}_n^c = \{(g,j) : \beta_{gj}^0 = 0\}$. In words, $\mathcal{A}_{g}$ is the active set for the $g$-th group, $\mathcal{A}_{g}^c$ is the non-active set for the $g$-th group, $\mathcal{A}_n$ is the active set across all groups, and $\mathcal{A}_n^c$ is the non-active set across all groups.

$$\pi\big(\boldsymbol{\beta}_n \ \big| \ \mathcal{G}_n, \mathcal{H}_n, \tau_n^2\big) = \prod_{g=1}^{G_n}\prod_{j=1}^{p_{g}} \frac{\Gamma(b_{g}+1/2)}{\Gamma(b_{g})\sqrt{2{\pi}\tau_n^2\gamma_{g}^2}}\bigg(1 + \frac{\beta_{gj}^2}{2\tau_n^2\gamma_{g}^2}\bigg)^{-(b_{g}+1/2)}$$

\noindent Then, we see that

$$\pi\bigg(\boldsymbol{\beta}_n : \Vert\boldsymbol{\beta}_n-\boldsymbol{\beta}_n^0\Vert_2 < \frac{\Delta}{n^{\rho/2}} \ \bigg| \ \mathcal{G}_n, \mathcal{H}_n, \tau_n^2\bigg)$$

$$\geq \prod_{g=1}^{G_n}\Bigg[\prod_{j \in \mathcal{A}_{g}}\pi\bigg(|\beta_{gj}-\beta_{gj}^0| < \frac{\Delta}{\sqrt{p_n}n^{\rho/2}} \ \bigg| \ \mathcal{G}_n, \mathcal{H}_n, \tau_n^2 \bigg)\times\prod_{j \in \mathcal{A}_{g}^c}\pi\bigg(|\beta_{gj}| < \frac{\Delta}{\sqrt{p_n}n^{\rho/2}} \ \bigg| \ \mathcal{G}_n, \mathcal{H}_n, \tau_n^2\bigg)\Bigg].$$

\noindent Continuing

$$\pi\bigg(|\beta_{gj}-\beta_{gj}^0| < \frac{\Delta}{\sqrt{p_n}n^{\rho/2}} \ \bigg| \ \mathcal{G}_n, \mathcal{H}_n, \tau_n^2 \bigg)$$ $$= \int_{\beta_{gj}^0 - \frac{\Delta}{\sqrt{p_n}n^{\rho/2}}}^{\beta_{gj}^0 + \frac{\Delta}{\sqrt{p_n}n^{\rho/2}}} \frac{\Gamma(b_{g}+1/2)}{\Gamma(b_{g})\sqrt{2{\pi}\tau_n^2\gamma_{g}^2}}\bigg(1 + \frac{\beta_{gj}^2}{2\tau_n^2\gamma_{g}^2}\bigg)^{-(b_{g}+1/2)}d\beta_{gj}$$ $$\geq \frac{2\Delta\Gamma(b_{g}+1/2)}{\Gamma(b_{g})\sqrt{2{\pi}\tau_n^2\gamma_{g}^2}\sqrt{p_n}n^{\rho/2}}\Bigg(1 + \frac{(L_n + \frac{\Delta}{\sqrt{p_n}n^{\rho/2}})^2}{2\tau_n^2\gamma_{g}^2}\Bigg)^{-(b_{g}+1/2)}$$

\noindent and

$$\pi\bigg(|\beta_{gj}| < \frac{\Delta}{\sqrt{p_n}n^{\rho/2}} \ \bigg| \ \mathcal{G}_n, \mathcal{H}_n, \tau_n^2 \bigg)$$ $$= \int_{-\frac{\Delta}{\sqrt{p_n}n^{\rho/2}}}^{\frac{\Delta}{\sqrt{p_n}n^{\rho/2}}} \frac{\Gamma(b_{g}+1/2)}{\Gamma(b_{g})\sqrt{2{\pi}\tau_n^2\gamma_{g}^2}}\bigg(1 + \frac{\beta_{gj}^2}{2\tau_n^2\gamma_{g}^2}\bigg)^{-(b_{g}+1/2)}d\beta_{gj}$$ $$= 2\int_{0}^{\frac{\Delta}{\sqrt{p_n}n^{\rho/2}}} \frac{\Gamma(b_{g}+1/2)}{\Gamma(b_{g})\sqrt{2{\pi}\tau_n^2\gamma_{g}^2}}\bigg(1+\frac{\beta_{gj}^2}{2\tau_n^2\gamma_{g}^2}\bigg)^{-(b_{g}+1/2)}d\beta_{gj}$$ $$\geq 2\int_{0}^{\frac{\Delta}{\sqrt{p_n}n^{\rho/2}}} \frac{\Gamma(b_{g}+1/2)}{\Gamma(b_{g})\sqrt{2{\pi}\tau_n^2\gamma_{g}^2}}\exp\Bigg(-\frac{\beta_{gj}(b_{g}+1/2)}{\sqrt{2\tau_n^2\gamma_{g}^2}}\Bigg)d\beta_{gj}$$ $$= \frac{2\Gamma(b_{g}+1/2)}{\Gamma(b_{g})\sqrt{2{\pi}\tau_n^2\gamma_{g}^2}}\Bigg(\frac{\sqrt{2\tau_n^2\gamma_{g}^2}}{b_{g}+1/2}\Bigg)\Bigg[1-\exp\Bigg(-\frac{\Delta(b_{g}+1/2)}{\sqrt{2\tau_n^2\gamma_{g}^2p_nn^{\rho}}}\Bigg)\Bigg]$$ $$= \frac{2\Gamma(b_{g}+1/2)}{\Gamma(b_{g})\sqrt{\pi}(b_{g}+1/2)}\Bigg[1-\exp\Bigg(-\frac{\Delta(b_{g}+1/2)}{\sqrt{2\tau_n^2\gamma_{g}^2p_nn^{\rho}}}\Bigg)\Bigg].$$

\noindent Note that we have the above inequality because $(1+x)^{-1} \geq \exp(-\sqrt{x})$ for all $x \geq 0$. Therefore,

$$\prod_{g=1}^{G_n}\Bigg[\prod_{j \in \mathcal{A}_{g}}\pi\bigg(|\beta_{gj}-\beta_{gj}^0| < \frac{\Delta}{\sqrt{p_n}n^{\rho/2}} \ \bigg| \ \mathcal{G}_n, \mathcal{H}_n, \tau_n^2 \bigg)\times\prod_{j \in \mathcal{A}_{g}^c}\pi\bigg(|\beta_{gj}| < \frac{\Delta}{\sqrt{p_n}n^{\rho/2}} \ \bigg| \ \mathcal{G}_n, \mathcal{H}_n, \tau_n^2\bigg)\Bigg]$$ $$\geq \prod_{g=1}^{G_n}\Bigg(\frac{2\Delta\Gamma(b_{g}+1/2)}{\Gamma(b_{g})\sqrt{2{\pi}\tau_n^2\gamma_{g}^2}\sqrt{p_n}n^{\rho/2}}\Bigg)^{|\mathcal{A}_{g}|}\Bigg(1 + \frac{(L_n + \frac{\Delta}{\sqrt{p_n}n^{\rho/2}})^2}{2\tau_n^2\gamma_{g}^2}\Bigg)^{-|\mathcal{A}_{g}|(b_{g}+1/2)}$$ $$\times \Bigg(\frac{2\Gamma(b_{g}+1/2)}{\Gamma(b_{g})\sqrt{\pi}(b_{g}+1/2)}\Bigg)^{|\mathcal{A}_{g}^c|}\Bigg[1-\exp\Bigg(-\frac{\Delta(b_{g}+1/2)}{\sqrt{2\tau_n^2\gamma_{g}^2p_nn^{\rho}}}\Bigg)\Bigg]^{|\mathcal{A}_{g}^c|}.$$

\noindent Substituting in $\tau_n^2 = C/(p_nn^{\rho}\log(n))$ and taking the negative logarithm of the final expression yields

$$- \sum_{g=1}^{G_n}\Bigg[|\mathcal{A}_{g}|\log\Bigg(\frac{2\Delta\Gamma(b_{g}+1/2)\sqrt{\log(n)}}{\Gamma(b_{g})\sqrt{2C{\pi}\gamma_{g}^2}}\Bigg)$$ $$- |\mathcal{A}_{g}|(b_{g}+1/2)\log\Bigg(1 + \frac{p_nn^{\rho}\log(n)(L_n + \frac{\Delta}{\sqrt{p_n}n^{\rho/2}})^2}{2C\gamma_{g}^2}\Bigg)$$ $$+ |\mathcal{A}_{g}^c|\log\Bigg(\frac{2\Gamma(b_{g}+1/2)}{\Gamma(b_{g})\sqrt{\pi}(b_{g}+1/2)}\Bigg) + |\mathcal{A}_{g}^c|\log\Bigg(1-\exp\Bigg(-\frac{\Delta(b_{g}+1/2)\sqrt{\log(n)}}{\sqrt{2C\gamma_{g}^2}}\Bigg)\Bigg)\Bigg]$$

$$= \sum_{g=1}^{G_n}\Bigg[-|\mathcal{A}_{g}|\log\Bigg(\frac{2\Delta\Gamma(b_{g}+1/2)\sqrt{\log(n)}}{\Gamma(b_{g})\sqrt{2C{\pi}\gamma_{g}^2}}\Bigg)$$ $$+ |\mathcal{A}_{g}|(b_{g}+1/2)\log\Bigg(1 + \frac{p_nn^{\rho}\log(n)(L_n + \frac{\Delta}{\sqrt{p_n}n^{\rho/2}})^2}{2C\gamma_{g}^2}\Bigg)$$ $$- |\mathcal{A}_{g}^c|\log\Bigg(\frac{2\Gamma(b_{g}+1/2)}{\Gamma(b_{g})\sqrt{\pi}(b_{g}+1/2)}\Bigg) - |\mathcal{A}_{g}^c|\log\Bigg(1-\exp\Bigg(-\frac{\Delta(b_{g}+1/2)\sqrt{\log(n)}}{\sqrt{2C\gamma_{g}^2}}\Bigg)\Bigg)\Bigg].$$

\noindent Let $$T_{n1} = \inf_{g \in \{1,...,G_n\}} \log\Bigg(\frac{2\Delta\Gamma(b_{g}+1/2)}{\Gamma(b_{g})\sqrt{2C{\pi}\gamma_{g}^2}}\Bigg),$$ $$T_{n2} = \inf_{g \in \{1,...,G_n\}} \log\Bigg(\frac{2\Gamma(b_{g}+1/2)}{\Gamma(b_{g})\sqrt{\pi}(b_{g}+1/2)}\Bigg),$$ $$T_{n3} = \inf_{g \in \{1,...,G_n\}} \frac{\Delta(b_{g}+1/2)}{\sqrt{2C\gamma_{g}^2}},$$ $$\gamma_{n,min}^2 = \inf_{g \in \{1,...,G_n\}}\gamma_{g}^2, \hspace{2 mm} \text{and} \hspace{2 mm} b_{n,max} = \sup_{g \in \{1,...,G_n\}}b_{g}.$$

Then, $$\sum_{g=1}^{G_n}\Bigg[-|\mathcal{A}_{g}|\log\Bigg(\frac{2\Delta\Gamma(b_{g}+1/2)\sqrt{\log(n)}}{\Gamma(b_{g})\sqrt{2C{\pi}\gamma_{g}^2}}\Bigg)$$ $$+ |\mathcal{A}_{g}|(b_{g}+1/2)\log\Bigg(1 + \frac{p_nn^{\rho}\log(n)(L_n + \frac{\Delta}{\sqrt{p_n}n^{\rho/2}})^2}{2C\gamma_{g}^2}\Bigg)$$ $$- |\mathcal{A}_{g}^c|\log\Bigg(\frac{2\Gamma(b_{g}+1/2)}{\Gamma(b_{g})\sqrt{\pi}(b_{g}+1/2)}\Bigg) - |\mathcal{A}_{g}^c|\log\Bigg(1-\exp\Bigg(-\frac{\Delta(b_{g}+1/2)\sqrt{\log(n)}}{\sqrt{2C\gamma_{g}^2}}\Bigg)\Bigg)\Bigg]$$

$$\leq \sum_{g=1}^{G_n}\Bigg[-\frac{|\mathcal{A}_{g}|}{2}\log\big(\log(n)\big) - |\mathcal{A}_{g}|T_{n1} + |\mathcal{A}_{g}|(b_{n,max}+1/2)\log\Bigg(1 + \frac{p_nn^{\rho}\log(n)(L_n + \frac{\Delta}{\sqrt{p_n}n^{\rho/2}})^2}{2C\gamma_{n,min}^2}\Bigg)$$ $$- |\mathcal{A}_{g}^c|T_{n2} - |\mathcal{A}_{g}^c|\log\bigg(1 - \exp\Big(T_{n3}\sqrt{\log(n)}\Big)\bigg)\Bigg]$$

$$= -\frac{|\mathcal{A}_n|}{2}\log\big(\log(n)\big) - |\mathcal{A}_n|T_{n1} + |\mathcal{A}_n|(b_{n,max}+1/2)\log\Bigg(1 + \frac{p_nn^{\rho}\log(n)(L_n + \frac{\Delta}{\sqrt{p_n}n^{\rho/2}})^2}{2C\gamma_{n,min}^2}\Bigg)$$ $$- |\mathcal{A}_n^c|T_{n2} - |\mathcal{A}_n^c|\log\bigg(1 - \exp\Big(-T_{n3}\sqrt{\log(n)}\Big)\bigg).$$

\noindent Note that the above expression is dominated by the $$|\mathcal{A}_n|(b_{n,max}+1/2)\log\Bigg(1 + \frac{p_nn^{\rho}\log(n)(L_n + \frac{\Delta}{\sqrt{p_n}n^{\rho/2}})^2}{2C\gamma_{n,min}^2}\Bigg)$$ term. If $|\mathcal{A}_n| = o(n/\log(n))$, then by \textit{Theorem 1} in \cite{armagan2013}, we obtain posterior consistency conditional on $\mathcal{G}_n$, i.e., $\forall \epsilon > 0$ $$\pi\big(\boldsymbol{\beta}_n : \Vert\boldsymbol{\beta}_n-\boldsymbol{\beta}_n^0\Vert_2 < \epsilon \ \big| \ \boldsymbol{y}_n, \mathcal{G}_n, \mathcal{H}_n, \tau_n^2, \sigma^2\big) \to 1$$ almost surely.


Lastly, by combining the law of iterated expectations with the dominated convergence theorem, we obtain $$\pi\big(\boldsymbol{\beta}_n: \Vert\boldsymbol{\beta}_n-\boldsymbol{\beta}_n^0\Vert_2 < \epsilon \ \big| \ \boldsymbol{y}_n, \mathcal{H}_n, \tau_n^2, \sigma^2\big)$$ $$= E_{\mathcal{G}_n \mid \boldsymbol{y}_n, \mathcal{H}_n, \tau_n^2, \sigma^2}\bigg[\pi\big(\boldsymbol{\beta}_n : \Vert\boldsymbol{\beta}_n-\boldsymbol{\beta}_n^0\Vert_2 < \epsilon \ \big| \ \boldsymbol{y}_n, \mathcal{G}_n, \mathcal{H}_n, \tau_n^2, \sigma^2\big)\bigg]$$ $$\to E_{\mathcal{G}_n \mid \boldsymbol{y}_n, \mathcal{H}_n, \tau_n^2, \sigma^2}[1] = 1.$$

\newpage

\subsection{Proof of Theorem \ref{thm:concentration_regression}} \label{proof:concentration_regression}

Let $\boldsymbol{M} = \boldsymbol{I}_p-(\boldsymbol{I}_p + (\boldsymbol{X}^{\top}\boldsymbol{X})^{-1}\boldsymbol{D})^{-1} = (\boldsymbol{I}_p + \boldsymbol{D}^{-1}\boldsymbol{X}^{\top}\boldsymbol{X})^{-1}$ where $\boldsymbol{D}^{-1} = \sigma^{-2}\tau^2\boldsymbol{\Gamma}\boldsymbol{\Lambda}$. The Rayleigh quotient inequality yields, $$\Big\Vert\hat{\boldsymbol{\beta}}^{OLS}-E[\boldsymbol{\beta}\mid\cdot]\Big\Vert_2^2 \geq e_{min}(\boldsymbol{M}^{\top}\boldsymbol{M})\Big\Vert\hat{\boldsymbol{\beta}}^{OLS}\Big\Vert_2^2 = \big(\theta_{min}(\boldsymbol{M})\big)^2\Big\Vert\hat{\boldsymbol{\beta}}^{OLS}\Big\Vert_2^2,$$ where $e_{min}(\boldsymbol{M}^{\top}\boldsymbol{M})$ denotes the minimum eigenvalue of $\boldsymbol{M}^{\top}\boldsymbol{M}$ and $\theta_{min}(\boldsymbol{M})$ denotes the minimum singular value of $\boldsymbol{M}$. Then, we have $$\big(\theta_{min}(\boldsymbol{M})\big)^2\Big\Vert\hat{\boldsymbol{\beta}}^{OLS}\Big\Vert_2^2 = \frac{1}{\big(\theta_{max}(\boldsymbol{I}_p + \boldsymbol{D}^{-1}\boldsymbol{X}^{\top}\boldsymbol{X})\big)^2}\Big\Vert\hat{\boldsymbol{\beta}}^{OLS}\Big\Vert_2^2 = \frac{1}{\Vert\boldsymbol{I}_p + \boldsymbol{D}^{-1}\boldsymbol{X}^{\top}\boldsymbol{X}\Vert_{\mathcal{O}}^2}\Big\Vert\hat{\boldsymbol{\beta}}^{OLS}\Big\Vert_2^2,$$ where $\Vert\cdot\Vert_{\mathcal{O}}$ is the operator $2$-norm and $\Vert\cdot\Vert_2$ is the usual $L_2$-norm. The operator 2-norm is sub-additive and sub-multiplicative, implying that $$\frac{1}{\Vert\boldsymbol{I}_p + \boldsymbol{D}^{-1}\boldsymbol{X}^{\top}\boldsymbol{X}\Vert_{\mathcal{O}}^2}\Big\Vert\hat{\boldsymbol{\beta}}^{OLS}\Big\Vert_2^2 \geq \frac{1}{\big(\Vert\boldsymbol{I}_p\Vert_{\mathcal{O}} + \Vert\boldsymbol{D}^{-1}\Vert_{\mathcal{O}}\Vert\boldsymbol{X}^{\top}\boldsymbol{X}\Vert_{\mathcal{O}}\big)^2}\Big\Vert\hat{\boldsymbol{\beta}}^{OLS}\Big\Vert_2^2$$ $$= \bigg(\frac{1}{1+\theta_{max}(\boldsymbol{X}^{\top}\boldsymbol{X})\sigma^{-2}\tau^2\max_{(g,j)}\gamma_g^2\lambda_{gj}^2}\bigg)^2\Big\Vert\hat{\boldsymbol{\beta}}^{OLS}\Big\Vert_2^2$$

\noindent First, note that $$\pi\bigg(\frac{1}{1+c\sigma^{-2}\tau^2\max_{(g,j)}\gamma_g^2\lambda_{gj}^2} < \epsilon \ \bigg| \ \boldsymbol{y},\mathcal{H},\tau^2,\sigma^2\bigg) \leq \pi\bigg(\bigcup_{g=1}^{G}\bigcup_{j=1}^{p_g}\bigg\{\frac{1}{1+c\sigma^{-2}\tau^2\gamma_g^2\lambda_{gj}^{2}} < \epsilon\bigg\} \ \bigg| \ \boldsymbol{y},\mathcal{H},\tau^2,\sigma^2\bigg)$$ $$\leq \sum_{g=1}^{G}\sum_{j=1}^{p_g}\pi\bigg(\frac{1}{1+c\sigma^{-2}\tau^2\gamma_g^2\lambda_{gj}^{2}} < \epsilon \ \bigg| \ \boldsymbol{y},\mathcal{H},\tau^2,\sigma^2\bigg),$$ where $c = \theta_{max}(\boldsymbol{X}^{\top}\boldsymbol{X})$. Define $$K_{gj} = \frac{1}{1+c\sigma^{-2}\tau^2\gamma_g^2\lambda_{gj}^{2}}$$ and for notational simplicity let $\mathcal{K} = \mathcal{K}_{-g}\cup\{K_{gl}\}$ where $\mathcal{K}_{-g} = \{K_{g'1} : g' \neq g\}$. Also, let $\mathcal{L} = \cup_{g=1}^{G}\mathcal{L}_g$ denote the collection of all local shrinkage parameters where $\mathcal{L}_g = \{\lambda_{gj}^2 : 1 \leq j \leq p_g\}$. Then, $$\pi\big(K_{gl} < \epsilon \ \big| \ \boldsymbol{y},\mathcal{H},\tau^2,\sigma^2\big) = \frac{1}{\pi(\boldsymbol{y} \mid \mathcal{H},\tau^2,\sigma^2)}\int_{0}^{\epsilon} \pi(\boldsymbol{y}\mid K_{gl},\mathcal{H},\tau^2,\sigma^2)\pi(K_{gl}\mid\mathcal{H},\tau^2,\sigma^2)dK_{gl}$$ $$= \frac{1}{\pi(\boldsymbol{y}\mid\mathcal{H},\tau^2,\sigma^2)}\int_{0}^{\epsilon}\int_{(0,1)^{p-1}}\int_{(0,\infty)^p} \pi(\boldsymbol{y}\mid\mathcal{K},\mathcal{L},\mathcal{H},\tau^2,\sigma^2)\pi(\mathcal{K}\mid\mathcal{L},\mathcal{H},\tau^2,\sigma^2)$$ $$\times\pi(\mathcal{L}\mid\mathcal{H},\tau^2,\sigma^2)d\mathcal{L}d\mathcal{K}_{-g}dK_{gl},$$ where $(0,1)^{p-1}$ indicates a $p-1$ dimensional hypercube on $(0,1)$ and $(0,\infty)^p = (0,\infty)\times\cdots\times(0,\infty)$ $p$ times.
Looking at the individual components we first observe that $\pi(\boldsymbol{y}\mid\mathcal{K},\mathcal{L},\mathcal{H},\tau^2,\sigma^2)$ is just a reparameterized version of $\pi(\boldsymbol{y}\mid\boldsymbol{\Gamma},\boldsymbol{\Lambda},\mathcal{H},\tau^2,\sigma^2)$, where $$[\boldsymbol{y}\mid\boldsymbol{\Gamma},\boldsymbol{\Lambda},\mathcal{H},\tau^2,\sigma^2] \sim N(\boldsymbol{0}, \sigma^2\boldsymbol{I}_n+\tau^2\boldsymbol{X}\boldsymbol{\Gamma}\boldsymbol{\Lambda}\boldsymbol{X}^{\top}).$$ Note that $\tau^2\boldsymbol{X}\boldsymbol{\Gamma}\boldsymbol{\Lambda}\boldsymbol{X}^{\top}$ is a symmetric, positive definite matrix and therefore has positive, real eigenvalues. Thus the determinant of $\sigma^2\boldsymbol{I}_n+\tau^2\boldsymbol{X}\boldsymbol{\Gamma}\boldsymbol{\Lambda}\boldsymbol{X}^{\top}$ satisfies $|\sigma^2\boldsymbol{I}_n+\tau^2\boldsymbol{X}\boldsymbol{\Gamma}\boldsymbol{\Lambda}\boldsymbol{X}^{\top}| \geq \big(\sigma^2\big)^n$ and we get that $\pi(\boldsymbol{y}\mid\boldsymbol{\Gamma},\boldsymbol{\Lambda},\mathcal{H},\tau^2,\sigma^2) \leq (2\pi\sigma^2)^{-n/2}$. Therefore, we conclude that, $$\pi(\boldsymbol{y}\mid\mathcal{K},\mathcal{L},\mathcal{H},\tau^2,\sigma^2) \leq (2\pi\sigma^2)^{-n/2}.$$

\noindent One helpful observation is that $$\pi(\mathcal{K}\mid\mathcal{L},\mathcal{H},\tau^2,\sigma^2)\pi(\mathcal{L}\mid\mathcal{H},\tau^2,\sigma^2)$$ $$= \pi(K_{gl}\mid\lambda_{gl}^2,\mathcal{H},\tau^2,\sigma^2)\pi(\lambda_{gl}^2\mid\mathcal{H},\tau^2,\sigma^2)\prod_{g' \neq g}^{}\pi(K_{g'1}\mid\mathcal{L}_{g'},\mathcal{H},\tau^2,\sigma^2)\pi(\mathcal{L}_{g'}\mid\mathcal{H},\tau^2,\sigma^2).$$ Thus, we get a simplified upper bound $$\pi\big(K_{gl} < \epsilon \ \big| \ \boldsymbol{y},\mathcal{H},\tau^2,\sigma^2\big)$$ $$\leq \frac{1}{\pi(\boldsymbol{y}\mid\mathcal{H},\tau^2,\sigma^2)}\int_{0}^{\epsilon}\int_{0}^{\infty}(2\pi\sigma^2)^{-n/2}\pi(K_{gl}\mid\lambda_{gl}^2,\mathcal{H},\tau^2,\sigma^2)\pi(\lambda_{gl}^2\mid\mathcal{H},\tau^2,\sigma^2)d\lambda_{gl}^2dK_{gl}.$$

Note that because we are conditioning on the local shrinkage parameter then calculating $\pi(K_{gl}\mid\lambda_{gl}^2,\mathcal{H},\tau^2,\sigma^2)$ is a single variable transformation problem, where $$\pi(K_{gl}\mid\lambda_{gl}^2,\mathcal{H},\tau^2,\sigma^2) = \frac{1}{\Gamma(a_g)}\bigg(\frac{\sigma^2}{c\tau^2\lambda_{gl}^2}\bigg)^{a_g}(1-K_{gl})^{a_g-1}(K_{gl})^{-(1+a_g)}\exp\bigg(-\frac{\sigma^2(1-K_{gl})}{c\tau^2\lambda_{gl}^2K_{gl}}\bigg).$$

\noindent Moreover, $$\pi(\lambda_{gl}^2\mid\mathcal{H},\tau^2,\sigma^2) = \frac{1}{\Gamma(b_g)}(\lambda_{gl}^2)^{-b_g-1}\exp\bigg(-\frac{1}{\lambda_{gl}^2}\bigg),$$ is simply the prior on the local shrinkage parameters. Therefore, $$\pi\big(K_{gl} < \epsilon \ \big| \ \boldsymbol{y},\mathcal{H},\tau^2,\sigma^2\big)$$ $$\leq \frac{1}{(2\pi\sigma^2)^{n/2}\pi(\boldsymbol{y}\mid\mathcal{H},\tau^2,\sigma^2)}\int_{0}^{\epsilon}\int_{0}^{\infty} \pi(K_{gl}\mid\lambda_{gl}^2,\mathcal{H},\tau^2,\sigma^2)\pi(\lambda_{gl}^2\mid\mathcal{H},\tau^2,\sigma^2)d\lambda_{gl}^2dK_{gl}$$ $$= \frac{1}{(2\pi\sigma^2)^{n/2}\Gamma(a_g)\Gamma(b_g)\pi(\boldsymbol{y}\mid\mathcal{H},\tau^2,\sigma^2)}\bigg(\frac{\sigma^2}{c\tau^2}\bigg)^{a_g}\times$$ $$\int_{0}^{\epsilon}(1-K_{gl})^{a_g-1}(K_{gl})^{-(1+a_g)}\int_{0}^{\infty} \big(\lambda_{gl}^2\big)^{-(a_g+b_g)-1}\exp\bigg(-\bigg(1+\frac{\sigma^2(1-K_{gl})}{c\tau^2K_{gl}}\bigg)\Big/\lambda_{gl}^2\bigg) d\lambda_{gl}^2dK_{gl}$$ $$= \frac{\Gamma(a_g+b_g)}{(2\pi\sigma^2)^{n/2}\Gamma(a_g)\Gamma(b_g)\pi(\boldsymbol{y}\mid\mathcal{H},\tau^2,\sigma^2)}\bigg(\frac{\sigma^2}{c\tau^2}\bigg)^{a_g}$$ $$\times\int_{0}^{\epsilon}(1-K_{gl})^{a_g-1}(K_{gl})^{-(1+a_g)}\bigg(1+\frac{\sigma^2(1-K_{gl})}{c\tau^2K_{gl}}\bigg)^{-(a_g+b_g)}dK_{gl}$$ $$= \frac{\Gamma(a_g+b_g)}{(2\pi\sigma^2)^{n/2}\Gamma(a_g)\Gamma(b_g)\pi(\boldsymbol{y}\mid\mathcal{H},\tau^2,\sigma^2)}\bigg(\frac{\sigma^2}{c\tau^2}\bigg)^{a_g}$$ $$\times\int_{0}^{\epsilon}(1-K_{gl})^{a_g-1}(K_{gl})^{-(1+a_g)}\bigg(\frac{c\tau^2K_{gl}}{c\tau^2K_{gl}+\sigma^2(1-\epsilon)}\bigg)^{(a_g+b_g)}dK_{gl}$$ $$\leq \frac{\Gamma(a_g+b_g)(1-\epsilon)^{-(a_g+b_g)}}{(2\pi\sigma^2)^{n/2}\Gamma(a_g)\Gamma(b_g)\pi(\boldsymbol{y}\mid\mathcal{H},\tau^2,\sigma^2)}\bigg(\frac{c\tau^2}{\sigma^2}\bigg)^{b_g}\int_{0}^{\epsilon}(1-K_{gl})^{a_g-1}(K_{gl})^{b_g-1}dK_{gl}$$ $$\leq \frac{\Gamma(a_g+b_g)\epsilon^{b_g}(1-\epsilon)^{-(a_g+b_g)}}{(2\pi\sigma^2)^{n/2}\Gamma(a_g)b_g\Gamma(b_g)\pi(\boldsymbol{y}\mid\mathcal{H},\tau^2,\sigma^2)}\bigg(\frac{c\tau^2}{\sigma^2}\bigg)^{b_g}\max\{1, (1-\epsilon)^{a_g-1}\}.$$

\noindent Lastly, we see that $$\lim_{\tau^2 \to 0}\pi(\boldsymbol{y}\mid\mathcal{H},\tau^2,\sigma^2) = \lim_{\tau^2 \to 0}\int_{(0,\infty)^G}\int_{(0,\infty)^p}\pi(\boldsymbol{y}\mid\mathcal{G},\mathcal{L},\mathcal{H},\tau^2,\sigma^2)\pi(\mathcal{G}\mid\mathcal{H},\tau^2,\sigma^2)\pi(\mathcal{L}\mid\mathcal{H},\tau^2,\sigma^2)d\mathcal{L}d\mathcal{G}.$$ Since $\pi(\boldsymbol{y}\mid\mathcal{G},\mathcal{L},\mathcal{H},\tau^2,\sigma^2) = \pi(\boldsymbol{y}\mid\boldsymbol{\Gamma},\boldsymbol{\Lambda},\mathcal{H},\tau^2,\sigma^2) \leq (2\pi\sigma^2)^{-n/2}$, then $$\int_{(0,\infty)^G}\int_{(0,\infty)^p}\pi(\boldsymbol{y}\mid\mathcal{G},\mathcal{L},\mathcal{H},\tau^2,\sigma^2)\pi(\mathcal{G}\mid\mathcal{H},\tau^2,\sigma^2)\pi(\mathcal{L}\mid\mathcal{H},\tau^2,\sigma^2)d\mathcal{L}d\mathcal{G} \leq (2\pi\sigma^2)^{-n/2},$$ and by the dominated convergence theorem we have that $$\lim_{\tau^2 \to 0}\pi(\boldsymbol{y}\mid\mathcal{H},\tau^2,\sigma^2) = \frac{1}{(2\pi\sigma^2)^{n/2}}\exp\bigg(-\frac{1}{2\sigma^2}\boldsymbol{y}^{\top}\boldsymbol{y}\bigg).$$

\noindent So, we conclude that $$\pi\bigg(\frac{1}{1+c\sigma^{-2}\tau^2\max_{(g,j)}\gamma_g^2\lambda_{gj}^2} < \epsilon \ \bigg| \ \boldsymbol{y},\mathcal{H},\tau^2,\sigma^2\bigg) \leq \sum_{g=1}^{G}\sum_{j=1}^{p_g} \pi\big(K_{gj} < \epsilon \ \big| \ \boldsymbol{y},\mathcal{H},\tau^2,\sigma^2\big)$$ $$\leq \sum_{g=1}^{G}\frac{p_g\Gamma(a_g+b_g)\epsilon^{b_g}(1-\epsilon)^{-(a_g+b_g)}}{(2\pi\sigma^2)^{n/2}\Gamma(a_g)b_g\Gamma(b_g)\pi(\boldsymbol{y}\mid\mathcal{H},\tau^2,\sigma^2)}\bigg(\frac{c\tau^2}{\sigma^2}\bigg)^{b_g}\max\{1, (1-\epsilon)^{a_g-1}\},$$ where the upper bound goes to zero as $\tau^2 \to 0$. Therefore, we have shown that for fixed $\epsilon \in (0,1)$, that $$\pi\bigg(\frac{1}{1+\theta_{max}(\boldsymbol{X}^{\top}\boldsymbol{X})\sigma^{-2}\tau^2\max_{(g,j)}\gamma_g^2\lambda_{gj}^2} \geq \epsilon \ \bigg| \ \boldsymbol{y},\mathcal{H},\tau^2,\sigma^2\bigg) \to 1$$ as $\tau^2 \to 0$. Since, $$\Big\Vert\hat{\boldsymbol{\beta}}^{OLS}-E[\boldsymbol{\beta}\mid\cdot]\Big\Vert_2^2 \geq \bigg(\frac{1}{1+\theta_{max}(\boldsymbol{X}^{\top}\boldsymbol{X})\sigma^{-2}\tau^2\max_{(g,j)}\gamma_g^2\lambda_{gj}^2}\bigg)^2\Big\Vert\hat{\boldsymbol{\beta}}^{OLS}\Big\Vert_2^2,$$ then we must also have that $$\pi\bigg(\Big\Vert\hat{\boldsymbol{\beta}}^{OLS}-E[\boldsymbol{\beta}\mid\cdot]\Big\Vert_2 \geq \epsilon\Big\Vert\hat{\boldsymbol{\beta}}^{OLS}\Big\Vert_2 \ \bigg| \ \boldsymbol{y},\mathcal{H},\tau^2,\sigma^2\bigg) \to 1$$ as $\tau^2 \to 0$.

\newpage

\subsection{Proof of Corollary \ref{thm:concentration_regression_b}} \label{proof:concentration_regression_b}

If $\boldsymbol{X}_g^{\top}\boldsymbol{X}_{g'} = \boldsymbol{0}$ for all $g \neq g'$, then we have $$E[\boldsymbol{\beta}_g\mid\cdot] = \bigg(\boldsymbol{I}_{p_g}+(\boldsymbol{X}_g^{\top}\boldsymbol{X}_g)^{-1}\frac{\sigma^2\gamma_g^2}{\tau^2}\boldsymbol{\Lambda}_g^{-1}\bigg)^{-1}\hat{\boldsymbol{\beta}}_g^{OLS}, \hspace{2 mm} \hat{\boldsymbol{\beta}}_g^{OLS} = (\boldsymbol{X}_g^{\top}\boldsymbol{X}_g)^{-1}\boldsymbol{X}_g^{\top}\boldsymbol{y},$$ where $\boldsymbol{\Lambda}_g = \text{diag}\big(\lambda_{g1}^2,\ldots,\lambda_{gp_g}^2\big)$. Following a similar argument to the proof of Theorem 3.3, we arrive at $$\Big\Vert\hat{\boldsymbol{\beta}}_g^{OLS}-E[\boldsymbol{\beta}_g\mid\cdot]\Big\Vert_2^2 \geq \bigg(\frac{1}{1+\theta_{max}(\boldsymbol{X}_g^{\top}\boldsymbol{X}_g)\sigma^{-2}\tau^2\gamma_g^2\max_{j}\lambda_{gj}^2}\bigg)^2\Big\Vert\hat{\boldsymbol{\beta}}_g^{OLS}\Big\Vert_2^2,$$ Also, from the proof of Theorem 3.3, we have that $$\pi\bigg(\frac{1}{1+c\sigma^{-2}\tau^2\gamma_g^2\max_{j}\lambda_{gj}^2} < \epsilon \ \bigg| \ \boldsymbol{y},\mathcal{H},\tau^2,\sigma^2\bigg) \leq \sum_{j=1}^{p_g} \pi\big(K_{gj} < \epsilon \ \big| \ \boldsymbol{y},\mathcal{H},\tau^2,\sigma^2\big)$$ $$\leq \frac{p_g\Gamma(a_g+b_g)\epsilon^{b_g}(1-\epsilon)^{-(a_g+b_g)}}{(2\pi\sigma^2)^{n/2}\Gamma(a_g)b_g\Gamma(b_g)\pi(\boldsymbol{y}\mid\mathcal{H},\tau^2,\sigma^2)}\bigg(\frac{c\tau^2}{\sigma^2}\bigg)^{b_g}\max\{1, (1-\epsilon)^{a_g-1}\},$$

\noindent If, $a_g \in (0,1)$ and $$\frac{c\epsilon\tau^2}{\sigma^2(1-\epsilon)} < 1$$ then $$\frac{p_g\Gamma(a_g+b_g)\epsilon^{b_g}(1-\epsilon)^{-(a_g+b_g)}}{(2\pi\sigma^2)^{n/2}\Gamma(a_g)b_g\Gamma(b_g)\pi(\boldsymbol{y}\mid\mathcal{H},\tau^2,\sigma^2)}\bigg(\frac{c\tau^2}{\sigma^2}\bigg)^{b_g}\max\{1, (1-\epsilon)^{a_g-1}\} \to 0$$ as $b_g \to \infty$. Since, $$\Big\Vert\hat{\boldsymbol{\beta}}_g^{OLS}-E[\boldsymbol{\beta}_g\mid\cdot]\Big\Vert_2^2 \geq \bigg(\frac{1}{1+\theta_{max}(\boldsymbol{X}_g^{\top}\boldsymbol{X}_g)\sigma^{-2}\tau^2\gamma_g^2\max_{j}\lambda_{gj}^2}\bigg)^2\Big\Vert\hat{\boldsymbol{\beta}}_g^{OLS}\Big\Vert_2^2,$$ then we must also have that for all $\delta \in (0, \sigma^2/(\sigma^2+\theta_{max}(\boldsymbol{X}_g^{\top}\boldsymbol{X}_g)\tau^2))$ $$\pi\bigg(\Big\Vert\hat{\boldsymbol{\beta}}_g^{OLS}-E[\boldsymbol{\beta}_g\mid\cdot]\Big\Vert_2 \geq \delta\Big\Vert\hat{\boldsymbol{\beta}}_g^{OLS}\Big\Vert_2\ \bigg| \ \boldsymbol{y},\mathcal{H},\tau^2,\sigma^2\bigg) \to 1$$ as $b_g \to \infty$.

\newpage

\subsection{Proof of Theorem \ref{thm:post_concentrate}a} \label{proof:post_concentrate_a}

The posterior distribution of the shrinkage weights in the $g$-th group are given by

$$\pi\big(\boldsymbol{\kappa}_{g}\mid\boldsymbol{y}_{g},\tau^2,\sigma^2,a_g,b_g\big) \propto \Bigg(1 + \frac{\tau^2}{\sigma^2}\sum_{j=1}^{p_g}\frac{\kappa_{gj}}{1-\kappa_{gj}}\Bigg)^{-(a_g+p_gb_g)}\Bigg(\prod_{j=1}^{p_g}\kappa_{gj}^{b_g-1/2}(1-\kappa_{gj})^{-(b_g+1)}\exp\bigg(-\frac{y_{gj}^2}{2\sigma^2}\kappa_{gj}\bigg)\Bigg),$$ where $\boldsymbol{\kappa}_g = (\kappa_{g1},...,\kappa_{gp_g})$, $0 < \kappa_{gj} < 1$ for all $1 \leq j \leq p_g$, and $\boldsymbol{y}_g = (y_{g1},...,y_{gp_g})^{\top}$.

$$\pi(\kappa_{gl} > \psi \mid \boldsymbol{y}_g,\tau^2,\sigma^2,a_g,b_g) = \frac{A_{gl}}{B_g},$$

\noindent where $$A_{gl} = \int_{\psi}^{1}\int_{0}^{1}\cdots\int_{0}^{1}\Bigg(1 + \frac{\tau^2}{\sigma^2}\sum_{j=1}^{p_g}\frac{\kappa_{gj}}{1-\kappa_{gj}}\Bigg)^{-(a_g+p_gb_g)}$$ $$\times\Bigg(\prod_{j=1}^{p_g}\kappa_{gj}^{b_g-1/2}(1-\kappa_{gj})^{-(b_g+1)}\exp\bigg(-\frac{y_{gj}^2}{2\sigma^2}\kappa_{gj}\bigg)\Bigg)d\kappa_{g1}{\cdots}d\kappa_{g,l-1}d\kappa_{g,l+1}{\cdots}d\kappa_{gp_g}d\kappa_{gl}$$ and

$B_g =$ $$\int_{0}^{1}\cdots\int_{0}^{1}\Bigg(1 + \frac{\tau^2}{\sigma^2}\sum_{j=1}^{p_g}\frac{\kappa_{gj}}{1-\kappa_{gj}}\Bigg)^{-(a_g+p_gb_g)}\Bigg(\prod_{j=1}^{p_g}\kappa_{gj}^{b_g-1/2}(1-\kappa_{gj})^{-(b_g+1)}\exp\bigg(-\frac{y_{gj}^2}{2\sigma^2}\kappa_{gj}\bigg)\Bigg)d\kappa_{g1}{\cdots}d\kappa_{gp_g}.$$

\noindent Note that $$\bigg(1 + \frac{\tau^2}{\sigma^2}\sum_{j=1}^{p_g}\frac{\kappa_{gj}}{1-\kappa_{gj}}\bigg)^{-(a_g+p_gb_g)}$$ $$= \bigg(1 + \frac{\tau^2}{\sigma^2}\sum_{j=1}^{p_g}\frac{\kappa_{gj}}{1-\kappa_{gj}}\bigg)^{-(a_g/p_g+b_g)}\bigg(1 + \frac{\tau^2}{\sigma^2}\sum_{j=1}^{p_g}\frac{\kappa_{gj}}{1-\kappa_{gj}}\bigg)^{-(p_g-1)(a_g/p_g+b_g)}$$

$$\leq \bigg(1+\frac{\tau^2}{\sigma^2}\frac{\kappa_{gl}}{1-\kappa_{gl}}\bigg)^{-(a_g/p_g+b_g)}\bigg(1+\frac{\tau^2}{\sigma^2}\sum_{j \neq l}\frac{\kappa_{gj}}{1-\kappa_{gj}}\bigg)^{-(p_g-1)(a_g/p_g+b_g)}$$

\noindent Then,

$$A_{gl} \leq \Bigg(\int_{0}^{1}\cdots\int_{0}^{1}\bigg(1+\frac{\tau^2}{\sigma^2}\sum_{j \neq l}\frac{\kappa_{gj}}{1-\kappa_{gj}}\bigg)^{-(p_g-1)(a_g/p_g+b_g)}\prod_{j \neq l}\kappa_{gj}^{b_g-1/2}(1-\kappa_{gj})^{-(b_g+1)}\exp\bigg(-\frac{y_{gj}^2}{2\sigma^2}\kappa_{gj}\bigg)d\kappa_{gj}\Bigg) \times$$

$$\Bigg(\int_{\psi}^{1}\bigg(1+\frac{\tau^2}{\sigma^2}\frac{\kappa_{gl}}{1-\kappa_{gl}}\bigg)^{-(a_g/p_g+b_g)}\kappa_{gl}^{b_g-1/2}(1-\kappa_{gl})^{-(b_g+1)}\exp\bigg(-\frac{y_{gl}^2}{2\sigma^2}\kappa_{gl}\bigg)d\kappa_{gl}\Bigg)$$

$$\leq \Bigg(\int_{0}^{1}\cdots\int_{0}^{1}\bigg(1+\frac{\tau^2}{\sigma^2}\sum_{j \neq l}\frac{\kappa_{gj}}{1-\kappa_{gj}}\bigg)^{-(p_g-1)(a_g/p_g+b_g)}\prod_{j \neq l}\kappa_{gj}^{b_g-1/2}(1-\kappa_{gj})^{-(b_g+1)}d\kappa_{gj}\Bigg) \times$$

$$\exp\bigg(-\frac{\psi}{2\sigma^2}y_{gl}^2\bigg)\Bigg(\int_{\psi}^{1}\bigg(1+\frac{\tau^2}{\sigma^2}\frac{\kappa_{gl}}{1-\kappa_{gl}}\bigg)^{-(a_g/p_g+b_g)}\kappa_{gl}^{b_g-1/2}(1-\kappa_{gl})^{-(b_g+1)}d\kappa_{gl}\Bigg)$$

$$\leq \Bigg(\int_{0}^{1}\cdots\int_{0}^{1}\bigg(1+\frac{\tau^2}{\sigma^2}\sum_{j \neq l}\frac{\kappa_{gj}}{1-\kappa_{gj}}\bigg)^{-(a_g^{*}+(p_g-1)b_g)}\prod_{j \neq l}\kappa_{gj}^{b_g-1}(1-\kappa_{gj})^{-(b_g+1)}d\kappa_{gj}\Bigg) \times$$

$$\exp\bigg(-\frac{\psi}{2\sigma^2}y_{gl}^2\bigg)\Bigg(\int_{\psi}^{1}\bigg(1-\bigg(1-\frac{\tau^2}{\sigma^2}\bigg)\kappa_{gl}\bigg)^{-(a_g/p_g+b_g)}\kappa_{gl}^{b_g-1/2}(1-\kappa_{gl})^{a_g/p_g-1}d\kappa_{gl}\Bigg)$$

$$\leq \Bigg(\bigg(\frac{\tau^2}{\sigma^2}\bigg)^{-(p_g-1)b_g}\frac{\Gamma(a_g^{*})(\Gamma(b_g))^{p_g-1}}{\Gamma(a_g^{*}+(p_g-1)b_g)}\Bigg)\bigg(\min\bigg(1,\frac{\tau^2}{\sigma^2}\bigg)\bigg)^{-(a_g/p_g+b_g)}$$ $${\times}\exp\bigg(-\frac{\psi}{2\sigma^2}y_{gl}^2\bigg)\int_{\psi}^{1}\kappa_{gl}^{b_g-1/2}(1-\kappa_{gl})^{a_g/p_g-1}d\kappa_{gl}$$

$$\leq \Bigg(\bigg(\frac{\tau^2}{\sigma^2}\bigg)^{-(p_g-1)b_g}\frac{\Gamma(a_g^{*})(\Gamma(b_g))^{p_g-1}}{\Gamma(a_g^{*}+(p_g-1)b_g)}\Bigg)\bigg(\min\bigg(1,\frac{\tau^2}{\sigma^2}\bigg)\bigg)^{-(a_g/p_g+b_g)}\max\Big(1,\psi^{b_g-1/2}\Big)$$ $${\times}\exp\bigg(-\frac{\psi}{2\sigma^2}y_{gl}^2\bigg)\int_{\psi}^{1}(1-\kappa_{gl})^{a_g/p_g-1}d\kappa_{gl}$$

$$= \Bigg(\bigg(\frac{\tau^2}{\sigma^2}\bigg)^{-(p_g-1)b_g}\frac{\Gamma(a_g^{*})(\Gamma(b_g))^{p_g-1}}{\Gamma(a_g^{*}+(p_g-1)b_g)}\Bigg)\bigg(\min\bigg(1,\frac{\tau^2}{\sigma^2}\bigg)\bigg)^{-(a_g/p_g+b_g)}\max\Big(1,\psi^{b_g-1/2}\Big)$$ $${\times}\exp\bigg(-\frac{\psi}{2\sigma^2}y_{gl}^2\bigg)\frac{p_g}{a_g}(1-\psi)^{a_g/p_g},$$

\noindent where $a_g^{*} = (p_g-1)a_g/p_g$. We can simplify the integral four lines above based on the prior distribution of the shrinkage weights for a group of size $p_g-1$.

Next, let $\delta \in (0,1)$ be fixed constant. Then,

$$B_g \geq \int_{0}^{\psi\delta}\cdots\int_{0}^{\psi\delta}\Bigg(1 + \frac{\tau^2}{\sigma^2}\sum_{j=1}^{p_g}\frac{\kappa_{gj}}{1-\kappa_{gj}}\Bigg)^{-(a_g+p_gb_g)}\Bigg(\prod_{j=1}^{p_g}\kappa_{gj}^{b_g-1/2}(1-\kappa_{gj})^{-(b_g+1)}\exp\bigg(-\frac{y_{gj}^2}{2\sigma^2}\kappa_{gj}\bigg)d\kappa_{gj}\Bigg)$$

$$\geq \Bigg(1 + \frac{p_g\tau^2}{\sigma^2}\frac{\psi\delta}{1-\psi\delta}\Bigg)^{-(a_g+p_gb_g)}\prod_{j=1}^{p_g}\int_{0}^{\psi\delta}\kappa_{gj}^{b_g-1/2}(1-\kappa_{gj})^{-(b_g+1)}\exp\bigg(-\frac{y_{gj}^2}{2\sigma^2}\kappa_{gj}\bigg)d\kappa_{gj}$$

$$\geq \Bigg(1 + \frac{p_g\tau^2}{\sigma^2}\frac{\psi\delta}{1-\psi\delta}\Bigg)^{-(a_g+p_gb_g)}\exp\bigg(-\frac{\psi\delta}{2\sigma^2}\sum_{j=1}^{p_g}y_{gj}^2\bigg)\prod_{j=1}^{p_g}\int_{0}^{\psi\delta}\kappa_{gj}^{b_g-1/2}(1-\kappa_{gj})^{-(b_g+1)}d\kappa_{gj}$$

$$\geq \Bigg(1 + \frac{p_g\tau^2}{\sigma^2}\frac{\psi\delta}{1-\psi\delta}\Bigg)^{-(a_g+p_gb_g)}\exp\bigg(-\frac{\psi\delta}{2\sigma^2}\sum_{j=1}^{p_g}y_{gj}^2\bigg)\prod_{j=1}^{p_g}\int_{0}^{\psi\delta}\kappa_{gj}^{b_g-1/2}d\kappa_{gj}$$

$$= \Bigg(1 + \frac{p_g\tau^2}{\sigma^2}\frac{\psi\delta}{1-\psi\delta}\Bigg)^{-(a_g+p_gb_g)}\exp\bigg(-\frac{\psi\delta}{2\sigma^2}\sum_{j=1}^{p_g}y_{gj}^2\bigg)(b_g+1/2)^{-p_g}(\psi\delta)^{p_g(b_g+1/2)}.$$

\noindent Therefore, $$\frac{A_{gl}}{B_g} \leq \frac{f(p_g,\tau^2,\sigma^2,a_g,b_g,\psi)}{g(p_g,\tau^2,\sigma^2,a_g,b_g,\psi,\delta)}\exp\bigg(\frac{\psi\delta}{2\sigma^2}\sum_{j \neq l}y_{gj}^2\bigg)\exp\bigg(-\frac{\psi(1-\delta)}{2\sigma^2}y_{gl}^2\bigg),$$

\noindent where $$f(p_g,\tau^2,\sigma^2,a_g,b_g,\psi) = \Bigg(\bigg(\frac{\tau^2}{\sigma^2}\bigg)^{-(p_g-1)b_g}\frac{\Gamma(a_g^{*})(\Gamma(b_g))^{p_g-1}}{\Gamma(a_g^{*}+(p_g-1)b_g)}\Bigg)$$ $$\times\bigg(\min\bigg(1,\frac{\tau^2}{\sigma^2}\bigg)\bigg)^{-(a_g/p_g+b_g)}\max\Big(1,\psi^{b_g-1/2}\Big)\frac{p_g}{a_g}(1-\psi)^{a_g/p_g}$$ and $$g(p_g,\tau^2,\sigma^2,a_g,b_g,\psi,\delta) = \Bigg(1 + \frac{p_g\tau^2}{\sigma^2}\frac{\psi\delta}{1-\psi\delta}\Bigg)^{-(a_g+p_gb_g)}(b_g+1/2)^{-p_g}(\psi\delta)^{p_g(b_g+1/2)}.$$

If we take the limit of this upper bound as $|y_{gl}| \to \infty$, then we see that $\pi(\kappa_{gl} > \psi \mid \boldsymbol{y}_g,\tau^2,\sigma^2,a_g,b_g) \to 0$. This concludes the proof.

\newpage

\subsection{Proof of Theorem \ref{thm:post_concentrate}b} \label{proof:post_concentrate_b}

The posterior distribution of the shrinkage weights in the $g$-th group are given by

$$\pi\big(\boldsymbol{\kappa}_{g}\mid\boldsymbol{y}_{g},\tau^2,\sigma^2,a_g,b_g\big)$$ $$\propto \Bigg(1 + \frac{\tau^2}{\sigma^2}\sum_{j=1}^{p_g}\frac{\kappa_{gj}}{1-\kappa_{gj}}\Bigg)^{-(a_g+p_gb_g)}\Bigg(\prod_{j=1}^{p_g}\kappa_{gj}^{b_g-1/2}(1-\kappa_{gj})^{-(b_g+1)}\exp\bigg(-\frac{y_{gj}^2}{2\sigma^2}\kappa_{gj}\bigg)\Bigg),$$ where $\boldsymbol{\kappa}_g = (\kappa_{g1},...,\kappa_{gp_g})$, $0 < \kappa_{gj} < 1$ for all $1 \leq j \leq p_g$, and $\boldsymbol{y}_g = (y_{g1},...,y_{gp_g})^{\top}$.

$$\pi(\kappa_{gl} < \epsilon\mid\boldsymbol{y}_g,\tau^2,\sigma^2,a_g,b_g) = \frac{A_{gl}}{B_g},$$ where $$A_{gl} = \int_{0}^{\epsilon}\int_{0}^{1}\cdots\int_{0}^{1}\Bigg(1 + \frac{\tau^2}{\sigma^2}\sum_{j=1}^{p_g}\frac{\kappa_{gj}}{1-\kappa_{gj}}\Bigg)^{-(a_g+p_gb_g)}$$ $$\times\Bigg(\prod_{j=1}^{p_g}\kappa_{gj}^{b_g-1/2}(1-\kappa_{gj})^{-(b_g+1)}\exp\bigg(-\frac{y_{gj}^2}{2\sigma^2}\kappa_{gj}\bigg)\Bigg)d\kappa_{g1}{\cdots}d\kappa_{g,l-1}d\kappa_{g,l+1}{\cdots}d\kappa_{gp_g}d\kappa_{gl}$$ and

$B_g =$ $$\int_{0}^{1}\cdots\int_{0}^{1}\Bigg(1 + \frac{\tau^2}{\sigma^2}\sum_{j=1}^{p_g}\frac{\kappa_{gj}}{1-\kappa_{gj}}\Bigg)^{-(a_g+p_gb_g)}\Bigg(\prod_{j=1}^{p_g}\kappa_{gj}^{b_g-1/2}(1-\kappa_{gj})^{-(b_g+1)}\exp\bigg(-\frac{y_{gj}^2}{2\sigma^2}\kappa_{gj}\bigg)\Bigg)d\kappa_{g1}{\cdots}d\kappa_{gp_g}.$$

\noindent Note that $$\bigg(1 + \frac{\tau^2}{\sigma^2}\sum_{j=1}^{p_g}\frac{\kappa_{gj}}{1-\kappa_{gj}}\bigg)^{-(a_g+p_gb_g)}$$ $$= \bigg(1 + \frac{\tau^2}{\sigma^2}\sum_{j=1}^{p_g}\frac{\kappa_{gj}}{1-\kappa_{gj}}\bigg)^{-(a_g/p_g+b_g)}\bigg(1 + \frac{\tau^2}{\sigma^2}\sum_{j=1}^{p_g}\frac{\kappa_{gj}}{1-\kappa_{gj}}\bigg)^{-(p_g-1)(a_g/p_g+b_g)}$$

$$\leq \bigg(1+\frac{\tau^2}{\sigma^2}\frac{\kappa_{gl}}{1-\kappa_{gl}}\bigg)^{-(a_g/p_g+b_g)}\bigg(1+\frac{\tau^2}{\sigma^2}\sum_{j \neq l}\frac{\kappa_{gj}}{1-\kappa_{gj}}\bigg)^{-(p_g-1)(a_g/p_g+b_g)}$$

\noindent Then,

$$A_{gl} \leq \Bigg(\int_{0}^{1}\cdots\int_{0}^{1}\bigg(1+\frac{\tau^2}{\sigma^2}\sum_{j \neq l}\frac{\kappa_{gj}}{1-\kappa_{gj}}\bigg)^{-(p_g-1)(a_g/p_g+b_g)}\prod_{j \neq l}\kappa_{gj}^{b_g-1/2}(1-\kappa_{gj})^{-(b_g+1)}\exp\bigg(-\frac{y_{gj}^2}{2\sigma^2}\kappa_{gj}\bigg)d\kappa_{gj}\Bigg) \times$$

$$\Bigg(\int_{0}^{\epsilon}\bigg(1+\frac{\tau^2}{\sigma^2}\frac{\kappa_{gl}}{1-\kappa_{gl}}\bigg)^{-(a_g/p_g+b_g)}\kappa_{gl}^{b_g-1/2}(1-\kappa_{gl})^{-(b_g+1)}\exp\bigg(-\frac{y_{gl}^2}{2\sigma^2}\kappa_{gl}\bigg)d\kappa_{gl}\Bigg)$$

$$\leq \Bigg(\int_{0}^{1}\cdots\int_{0}^{1}\bigg(1+\frac{\tau^2}{\sigma^2}\sum_{j \neq l}\frac{\kappa_{gj}}{1-\kappa_{gj}}\bigg)^{-(p_g-1)(a_g/p_g+b_g)}\prod_{j \neq l}\kappa_{gj}^{b_g-1/2}(1-\kappa_{gj})^{-(b_g+1)}d\kappa_{gj}\Bigg) \times$$

$$\Bigg((1-\epsilon)^{-(b_g+1)}\int_{0}^{\epsilon}\kappa_{gl}^{b_g-1/2}d\kappa_{gl}\Bigg)$$

$$\leq \frac{\epsilon^{b_g+1/2}}{(b_g+1/2)(1-\epsilon)^{b_g+1}}\int_{0}^{1}\cdots\int_{0}^{1}\bigg(1+\frac{\tau^2}{\sigma^2}\sum_{j \neq l}\frac{\kappa_{gj}}{1-\kappa_{gj}}\bigg)^{-(a_g^{*}+(p_g-1)b_g)}\prod_{j \neq l}\kappa_{gj}^{b_g-1}\kappa_{gj}^{1/2}(1-\kappa_{gj})^{-(b_g+1)}d\kappa_{gj}$$

$$\leq \frac{\epsilon^{b_g+1/2}}{(b_g+1/2)(1-\epsilon)^{b_g+1}}\int_{0}^{1}\cdots\int_{0}^{1}\bigg(1+\frac{\tau^2}{\sigma^2}\sum_{j \neq l}\frac{\kappa_{gj}}{1-\kappa_{gj}}\bigg)^{-(a_g^{*}+(p_g-1)b_g)}\prod_{j \neq l}\kappa_{gj}^{b_g-1}(1-\kappa_{gj})^{-(b_g+1)}d\kappa_{gj}$$

$$= \Bigg(\frac{\epsilon^{b_g+1/2}}{(b_g+1/2)(1-\epsilon)^{b_g+1}}\Bigg)\Bigg(\bigg(\frac{\tau^2}{\sigma^2}\bigg)^{-(p_g-1)b_g}\frac{\Gamma(a_g^{*})(\Gamma(b_g))^{p_g-1}}{\Gamma(a_g^{*}+(p_g-1)b_g)}\Bigg)$$

\noindent where $a_g^{*} = (p_g-1)a_g/p_g$. We have the last equality based on the prior distribution of the shrinkage weights for a group of size $p_g-1$. Next,

$$B_g \geq \exp\bigg(-\frac{1}{2\sigma^2}\sum_{j=1}^{p_g}y_{gj}^2\bigg)\int_{0}^{1}\cdots\int_{0}^{1}\bigg(1 + \frac{\tau^2}{\sigma^2}\sum_{j=1}^{p_g}\frac{\kappa_{gj}}{1-\kappa_{gj}}\bigg)^{-(a_g+p_gb_g)}\prod_{j=1}^{p_g}\kappa_{gj}^{b_g-1/2}(1-\kappa_{gj})^{-(b_g+1)}d\kappa_{gj}$$

$$= \exp\bigg(-\frac{1}{2\sigma^2}\sum_{j=1}^{p_g}y_{gj}^2\bigg)\int_{0}^{1}\cdots\int_{0}^{1}\bigg(1 + \frac{\tau^2}{\sigma^2}\sum_{j=1}^{p_g}\frac{\kappa_{gj}}{1-\kappa_{gj}}\bigg)^{-(a_g+p_gb_g^{*})}\bigg(1 + \frac{\tau^2}{\sigma^2}\sum_{j=1}^{p_g}\frac{\kappa_{gj}}{1-\kappa_{gj}}\bigg)^{p_g/2}$$ $$\times\prod_{j=1}^{p_g}\kappa_{gj}^{b_g^{*}-1}(1-\kappa_{gj})^{-(b_g^{*}+1)}(1-\kappa_{gj})^{1/2}d\kappa_{gj}$$

$$\geq \exp\bigg(-\frac{1}{2\sigma^2}\sum_{j=1}^{p_g}y_{gj}^2\bigg)\int_{0}^{1}\cdots\int_{0}^{1}\bigg(1 + \frac{\tau^2}{\sigma^2}\sum_{j=1}^{p_g}\frac{\kappa_{gj}}{1-\kappa_{gj}}\bigg)^{p_g/2}\bigg(\prod_{j=1}^{p_g}(1-\kappa_{gj})^{1/2}\bigg)$$ $$\times\bigg(1 + \frac{\tau^2}{\sigma^2}\sum_{j=1}^{p_g}\frac{\kappa_{gj}}{1-\kappa_{gj}}\bigg)^{-(a_g+p_gb_g^{*})}\prod_{j=1}^{p_g}\kappa_{gj}^{b_g^{*}-1}(1-\kappa_{gj})^{-(b_g^{*}+1)}d\kappa_{gj}$$

$$\geq \exp\bigg(-\frac{1}{2\sigma^2}\sum_{j=1}^{p_g}y_{gj}^2\bigg)\int_{0}^{1}\cdots\int_{0}^{1}\Bigg(\prod_{j=1}^{p_g}\bigg(1-\kappa_{gj} + \frac{\tau^2}{\sigma^2}\kappa_{gj}\bigg)\Bigg)^{1/2}\bigg(1 + \frac{\tau^2}{\sigma^2}\sum_{j=1}^{p_g}\frac{\kappa_{gj}}{1-\kappa_{gj}}\bigg)^{-(a_g+p_gb_g^{*})}$$ $$\times\prod_{j=1}^{p_g}\kappa_{gj}^{b_g^{*}-1}(1-\kappa_{gj})^{-(b_g^{*}+1)}d\kappa_{gj}$$

$$\geq \exp\bigg(-\frac{1}{2\sigma^2}\sum_{j=1}^{p_g}y_{gj}^2\bigg)\Bigg(\min\bigg(1,\frac{\tau^2}{\sigma^2}\bigg)\Bigg)^{p_g/2}\int_{0}^{1}\cdots\int_{0}^{1}\bigg(1 + \frac{\tau^2}{\sigma^2}\sum_{j=1}^{p_g}\frac{\kappa_{gj}}{1-\kappa_{gj}}\bigg)^{-(a_g+p_gb_g^{*})}$$ $$\times\prod_{j=1}^{p_g}\kappa_{gj}^{b_g^{*}-1}(1-\kappa_{gj})^{-(b_g^{*}+1)}d\kappa_{gj}$$

$$= \exp\bigg(-\frac{1}{2\sigma^2}\sum_{j=1}^{p_g}y_{gj}^2\bigg)\Bigg(\min\bigg(1,\frac{\tau^2}{\sigma^2}\bigg)\Bigg)^{p_g/2}\Bigg(\bigg(\frac{\tau^2}{\sigma^2}\bigg)^{-p_gb_g^{*}}\frac{\Gamma(a_g)(\Gamma(b_g^{*}))^{p_g}}{\Gamma(a_g+p_gb_g^{*})}\Bigg),$$

\noindent where $b_g^{*} = b_g + 1/2$. Similarly, we have the last equality based on the prior distribution of the shrinkage weights for a group of size $p_g$.

Therefore, $$\frac{A_{gl}}{B_g} \leq \exp\bigg(\frac{1}{2\sigma^2}\sum_{j=1}^{p_g}y_{gj}^2\bigg)\frac{\epsilon^{b_g+1/2}}{(b_g+1/2)(1-\epsilon)^{b_g+1}}\bigg(\frac{\tau^2}{\sigma^2}\bigg)^{p_g/2+b_g}\Bigg(\min\bigg(1,\frac{\tau^2}{\sigma^2}\bigg)\Bigg)^{-p_g/2}$$ $$\times\frac{\Gamma(a_g+p_gb_g^{*})\Gamma(a_g^{*})(\Gamma(b_g))^{p_g-1}}{\Gamma(a_g^{*}+(p_g-1)b_g)\Gamma(a_g)(\Gamma(b_g^{*}))^{p_g}}.$$

If we take the limit of this expression as $\tau \to 0$, then we see that $\pi(\kappa_{gl} < \epsilon\mid\boldsymbol{y}_g,\tau^2,\sigma^2,a_g,b_g) \to 0$. This concludes the proof.

\newpage

\subsection{Proof of Corollary \ref{thm:post_group_shrink}} \label{proof:post_group_shrink}

From the proof of Theorem 3.4b we have that $$\pi(\kappa_{gl} < \epsilon\mid\boldsymbol{y}_g, \tau^2,\sigma^2,a_g,b_g)$$ $$\leq \exp\bigg(\frac{1}{2\sigma^2}\boldsymbol{y}_g^{\top}\boldsymbol{y}_g\bigg)\frac{\epsilon^{b_g+1/2}}{(b_g+1/2)(1-\epsilon)^{b_g+1}}\bigg(\frac{\tau^2}{\sigma^2}\bigg)^{p_g/2+b_g}\Bigg(\min\bigg(1,\frac{\tau^2}{\sigma^2}\bigg)\Bigg)^{-p_g/2}$$ $$\times\frac{\Gamma(a_g+p_gb_g^{*})\Gamma(a_g^{*})(\Gamma(b_g))^{p_g-1}}{\Gamma(a_g^{*}+(p_g-1)b_g)\Gamma(a_g)(\Gamma(b_g^{*}))^{p_g}},$$ where $p_g$ is the number of observations in group $g$, $b_g^* = b_g + 1/2$, and $a_g^* = (p_g-1)a_g/p_g$. Based on this inequality we just need to find for what values of $\theta$ the limit as $b_g \to \infty$ of $$\frac{\theta^{b_g}}{(b_g+1/2)}\frac{\Gamma(a_g+p_gb_g^{*})(\Gamma(b_g))^{p_g-1}}{\Gamma(a_g^{*}+(p_g-1)b_g)(\Gamma(b_g^{*}))^{p_g}}, \hspace{4 mm} \theta = \frac{\epsilon}{1-\epsilon}\frac{\tau^2}{\sigma^2}$$ goes to zero. To do so we will first need one useful asymptotic approximation (a consequence of 6.1.39 in \cite{abramowitz1972}): $$\lim_{x \to \infty}\frac{\Gamma(x+c)}{\Gamma(x)x^c} = 1,$$ for any $c \in \mathbb{R}$. Therefore, $$\frac{\theta^{b_g}}{(b_g+1/2)}\frac{\Gamma(a_g+p_gb_g^{*})(\Gamma(b_g))^{p_g-1}}{\Gamma(a_g^{*}+(p_g-1)b_g)(\Gamma(b_g^{*}))^{p_g}}$$ $$\sim \frac{\theta^{b_g}}{(b_g+1/2)\Gamma(b_g+1/2)}\frac{\Gamma(p_gb_g)(p_gb_g)^{a_g+p_g/2}(\Gamma(b_g))^{p_g-1}}{\Gamma((p_g-1)b_g)((p_g-1)b_g)^{a_g^*}(\Gamma(b_g))^{p_g-1}b_g^{(p_g-1)/2}}$$ $$= \frac{\theta^{b_g}}{\Gamma(b_g+3/2)}\frac{\Gamma(p_gb_g)p_g^{a_g+p_g/2}b_g^{a_g/p_g+1/2}}{\Gamma((p_g-1)b_g)(p_g-1)^{a_g^*}} \sim \frac{p_g^{a_g+p_g/2}}{(p_g-1)^{a_g^*}}\frac{\theta^{b_g}\Gamma(p_gb_g)b_g^{a_g/p_g+1/2}}{\Gamma(b_g)b_g^{3/2}\Gamma((p_g-1)b_g)}$$ $$= \frac{p_g^{a_g+p_g/2}}{(p_g-1)^{a_g^*}}\frac{\theta^{b_g}\Gamma(p_gb_g)b_g^{a_g/p_g-1}}{\Gamma(b_g)\Gamma((p_g-1)b_g)} = \frac{p_g^{a_g+p_g/2}}{(p_g-1)^{a_g^*}}\frac{\theta^{b_g}b_g^{a_g/p_g-1}}{\mathcal{B}(b_g,(p_g-1)b_g)}.$$ Stirling's approximation for the gamma function (6.1.39 in \cite{abramowitz1972}) can be used to get the following asymptotic approximation for the beta function, $$\mathcal{B}(x,y) \sim \sqrt{2\pi}\frac{x^{x-1/2}y^{y-1/2}}{(x+y)^{x+y-1/2}.}$$ Then, $$\frac{p_g^{a_g+p_g/2}}{(p_g-1)^{a_g^*}}\frac{\theta^{b_g}b_g^{a_g/p_g-1}}{\mathcal{B}(b_g,(p_g-1)b_g)} \sim \frac{p_g^{a_g+p_g/2}}{\sqrt{2\pi}(p_g-1)^{a_g^*}}\frac{\theta^{b_g}b_g^{a_g/p_g-1}(p_gb_g)^{p_gb_g-1/2}}{b_g^{b_g-1/2}((p_g-1)b_g)^{(p_g-1)b_g-1/2}}.$$ If we set $\theta = p_g^{-p_g}$, then this quantity becomes $$\frac{p_g^{a_g+p_g/2}}{\sqrt{2\pi}(p_g-1)^{a_g^*}}\frac{\theta^{b_g}b_g^{a_g/p_g-1}(p_gb_g)^{p_gb_g-1/2}}{b_g^{b_g-1/2}((p_g-1)b_g)^{(p_g-1)b_g-1/2}} = \frac{p_g^{a_g+p_g/2}}{\sqrt{2\pi}(p_g-1)^{a_g^*}}\frac{b_g^{a_g/p_g-1}b_g^{p_gb_g}(p_gb_g)^{-1/2}}{b_g^{b_g-1/2}((p_g-1)b_g)^{(p_g-1)b_g-1/2}}$$ $$= \frac{p_g^{a_g+(p_g-1)/2}}{\sqrt{2\pi}(p_g-1)^{a_g^*}}\frac{b_g^{a_g/p_g-1}b_g^{(p_g-1)b_g}b_g^{b_g}}{b_g^{b_g}((p_g-1)b_g)^{(p_g-1)b_g-1/2}} = \frac{p_g^{a_g+(p_g-1)/2}}{\sqrt{2\pi}(p_g-1)^{a_g^*}}\frac{b_g^{a_g/p_g-1}}{(p_g-1)^{(p_g-1)b_g}((p_g-1)b_g)^{-1/2}}$$ $$= \frac{p_g^{a_g+(p_g-1)/2}}{\sqrt{2\pi}(p_g-1)^{a_g^*-1/2}}\frac{b_g^{a_g/p_g-1/2}}{(p_g-1)^{(p_g-1)b_g}}$$ If $a_g \in (0,1)$, then this quantity converges to zero as $b_g \to \infty$.

To summarize the result, we have shown that if $\tau^2$, $\sigma^2$, $p_g \geq 2$ and $a_g \in (0,1)$ are all fixed, then there exists a constant $$\epsilon(\tau^2,\sigma^2,p_g) = \bigg(1+\frac{\tau^2}{\sigma^2}p_g^{p_g}\bigg)^{-1},$$ such that $\pi(\kappa_{gl} < \epsilon(\tau^2,\sigma^2,p_g)\mid\boldsymbol{y}_g, \tau^2,\sigma^2,a_g,b_g) \to 0$ as $b_g \to \infty$.

\newpage

\subsection{Full Conditional Distributions for Gibbs Sampler} \label{app:gibbs_sampler}

\noindent The full conditional distributions for all model parameters are $$[\boldsymbol{\alpha}\mid\cdot] \sim N\bigg(\big(\boldsymbol{C}^{\top}\boldsymbol{C}\big)^{-1}\boldsymbol{C}^{\top}(\boldsymbol{y}-\boldsymbol{X}\boldsymbol{\beta}),\sigma^2\big(\boldsymbol{C}^{\top}\boldsymbol{C}\big)^{-1}\bigg)$$ $$[\boldsymbol{\beta}\mid\cdot] \sim N\Bigg(\boldsymbol{Q}^{-1}\frac{1}{\sigma^2}\boldsymbol{X}^{\top}\Big(\boldsymbol{y} - \boldsymbol{C}\boldsymbol{\alpha}\Big),\boldsymbol{Q}^{-1}\Bigg), \hspace{2 mm} \boldsymbol{Q} = \frac{1}{\sigma^2}\boldsymbol{X}^{\top}\boldsymbol{X}+\frac{1}{\tau^2}\boldsymbol{\Gamma}^{-1}\boldsymbol{\Lambda}^{-1}$$ $$[\tau^2\mid\cdot] \sim IG\Bigg(\frac{p+1}{2},\frac{1}{2}\boldsymbol{\beta}^{\top}\boldsymbol{\Gamma}^{-1}\boldsymbol{\Lambda}^{-1}\boldsymbol{\beta}+\frac{1}{\nu}\Bigg), \hspace{2 mm} [\nu\mid\cdot] \sim IG\Bigg(1,\frac{1}{\tau^2}+\frac{1}{\sigma^2}\Bigg)$$ $$[\sigma^2\mid\cdot] \sim IG\Bigg(\frac{n+1}{2},\frac{1}{2}\big(\boldsymbol{y} - \boldsymbol{C}\boldsymbol{\alpha} - \boldsymbol{X}\boldsymbol{\beta}\big)^{\top}\big(\boldsymbol{y} - \boldsymbol{C}\boldsymbol{\alpha} - \boldsymbol{X}\boldsymbol{\beta}\big)+\frac{1}{\nu}\Bigg)$$ $$[\lambda_{gj}^2\mid\cdot] \sim IG\Bigg(b_g+\frac{1}{2},1+\frac{\beta_{gj}^2}{2\tau^2\gamma_g^2}\Bigg), \hspace{2 mm} [\gamma_g^{-2}\mid\cdot] \sim GIG\Bigg(\frac{p_g}{2}-a_g,\frac{1}{\tau^2}\sum_{j=1}^{p_g}\frac{\beta_{gj}^{2}}{\lambda_{gj}^{2}},2\Bigg),$$

\noindent where GIG refers to the generalized inverse Gaussian distribution \citep{hormann2014}.

\newpage

\begin{figure}[!ht]
    \centering
    \includegraphics[scale=1.0, height = 0.75\textheight, width = 1.0\linewidth]{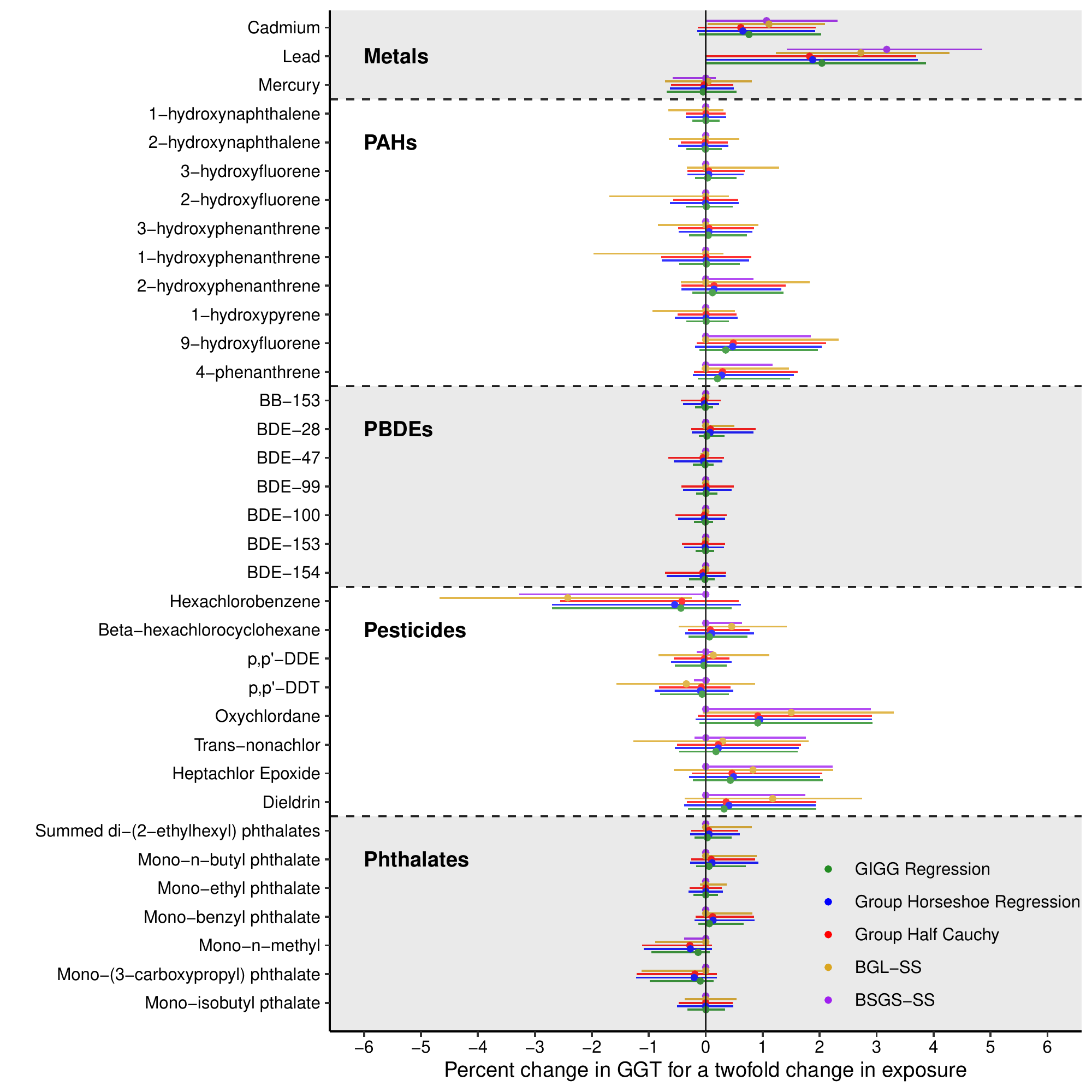}
    \caption{Associations between environmental toxicants (metals, phthalates, pesticides, PBDEs, and PAHs) and gamma glutamyl transferase (GGT) from NHANES 2003-2004 ($n = 990$).}
    \label{fig:ewas_forest_plot_glucose_supp}
\end{figure}

\end{document}